\documentclass[epj]{svjour}
\usepackage{epsf}
\begin{document}
\date{}
\title{Persistent currents in two dimensions: \\
New regimes induced by the interplay between 
electronic correlations and disorder}
\titlerunning{Persistent currents}

\author{Zolt\'an \'Ad\'am N\'emeth$^{(a,b)}$ and Jean-Louis 
Pichard$^{(a,c)}$}

\authorrunning{Z. \'A. N\'emeth and J.-L. Pichard}
        
\institute{
(a) CEA/DSM, Service de Physique de l'Etat Condens\'e, 
Centre d'Etudes de Saclay, 91191 Gif-sur-Yvette Cedex, France \\  
(b) E\"otv\"os University, Departement of Physics of Complex Systems, 
1117 Budapest, P\'azm\'any P\'eter s\'et\'any 1/A, Hungary \\
(c) Laboratoire de Physique Th\'eorique et Mod\'elisation, 
Universit\'e de Cergy-Pontoise, 95031, Cergy-Pontoise Cedex, France 
}

\abstract{Using the persistent current $I$ induced by an 
Aharonov-Bohm flux in square lattices with random potentials, 
we study the interplay between electronic correlations and 
disorder upon the ground state (GS) of a few polarized electrons 
(spinless fermions) with Coulomb repulsion. ${\bf K}$ being the 
total momentum, we show that $I \propto {\bf K}$ in the continuum 
limit. We use this relation to distinguish between the continuum 
regimes, where the lattice GS behaves as in the continuum limit and 
$I$ is independent of the interaction strength $U$ when ${\bf K}$ is 
conserved, and the lattice regimes where $I$ decays as $U$ increases. 
Changing the disorder strength $W$ and $U$, we obtain many regimes 
which we study using the map of local currents carried by three 
spinless fermions. The decays of $I$ characterizing three different 
lattice regimes are described by large $U$ perturbative expansions. 
In one of them, $I$ forms a stripe of current flowing along 
the axis of the diamagnetic Wigner molecule induced by large electronic 
correlations. This stripe of current persists in the continuum limit. 
The quantum melting of the diamagnetic molecule gives rise to an 
intermediate ``supersolid'' regime where a paramagnetic correlated pair 
co-exists with a third particle, before the total melting. The concepts 
of stripe and of supersolid which we use to describe certain regimes 
exhibited by three spinless fermions are reminiscent of the observations 
and conjectures done in other fields at the thermodynamic limit (stripe 
for high-Tc cuprates, supersolid for Helium quantum solids).}

\maketitle

\section{Introduction}
\label{section1}

\subsection{Motivations}
\label{subsection1.1}

 Let us first give some reasons which have motivated this study of a few 
polarized electrons (spinless fermions) and of their persistent current 
in two dimensional disordered lattice models. In the last decade, the 
conductance of two dimensional electron gases (2DEGs) have been 
measured \cite{abrahams} at low temperatures as a function of their 
densities. Using 2DEGs created in Ga-As heterostructures or Si-MOSFETs 
at densities smaller than those used for studying the weak-localization 
corrections to the Drude-Boltzmann conductance, a ``two dimensional 
metal-insulator transition'' has been observed. When the temperature 
increases, the conductance does not change at a certain critical density, 
while it increases at smaller densities and decreases at larger densities. 
The increase corresponds to the insulating phase and the decay to the 
metallic phase. In Ga-As heterostructures or Si-MOSFETs, there is a 
more or less important disorder due to charged impurities and rough 
interfaces. In presence of disorder, a 2DEG can be insulating for two 
reasons: either because it forms a Fermi glass with Anderson 
localization at high densities or because it forms a pinned Wigner crystal 
at low densities. Since this transition is observed at low densities, 
it was suggested \cite{yoon} that the low density insulator would be a 
pinned Wigner crystal, which would give an unexpected metal when it melts 
above the critical density. While a 2d insulator is expected, the 
observation of a 2d metal remains unexplained. If one goes \cite{simmons} 
from the Fermi glass towards the pinned Wigner crystal via a 2d metal as 
the density decreases, the existence of a new metallic phase between two 
insulating phases of different nature becomes unavoidable. And indeed, 
exact numerical studies of small disordered clusters have shown 
\cite{bwp1,bwp2,avishai,sp} the trace of an intermediate regime consistent 
with this hypothesis. 

Subsequent numerical studies of similar systems have shown \cite{ksp,np} 
that an intermediate regime persists without disorder. The possibility of 
having some hybrid phase made of a quantum solid co-existing with 
delocalized defects, was suggested in refs. \cite{ksp,np}, in analogy with 
the ``supersolid'' proposed long ago by Andreev and Lifshitz 
for quantum Helium solids in three dimensions \cite{andreev-lifshitz,DKL}. 
Using a similar analogy with Helium physics, but assuming macroscopic 
inhomogeneities of the carrier density due to an inhomogeneous substrate 
potential, a qualitative explanation of the dependence of the resistance 
as a function of a parallel magnetic field and of the temperature has been 
discussed in Ref. \cite{spivak1}.   

For a simple 2DEG without a random substrate, to understand how the ground 
state (GS) changes when one varies the carrier density remains a challenge 
despite a few decades of efforts. The simplest model for a 2DEG is a 
continuum Hamiltonian $H_c$ having three parts: one body kinetic terms, 
a two body Coulomb repulsion and a one body potential describing the effect 
of the positive ions. The last potential is necessary to have a stable system 
of electrons because of Coulomb repulsions. However, if one takes periodic 
boundary conditions (BCs) and a uniform jellium for the positive ions, 
the corresponding potential becomes a constant term. The scale of the one 
body quantum effects is given by the Bohr radius $a_B=\hbar^2/me^2$, $e$ 
and $m$ being the electronic charge and mass. The strength of the Coulomb 
repulsion depends on the radius $a$ of a circle which encloses on the 
average one electron. Measuring the energies in rydbergs ($1 Ry = 
me^4/2\hbar^2$) and the lengths in units of $a$, the continuum 2DEG 
Hamiltonian for $N$ electrons reads
\begin{equation}
H_c=-\frac{1}{r_s^2} \sum_{i=1}^N \nabla_i^2 + \frac{2}{r_s}  
\sum_{1\leq i < j\leq N} \frac{1}{|{\bf r}_i-{\bf r}_j|} + const, 
\label{H-continuum}
\end{equation}
which only depends on a single scaling ratio
\begin{equation}
r_s=\frac{a}{a_B}
\end{equation}
when $N \rightarrow \infty$. We neglect the spin degrees of freedom, 
considering fully polarized electrons (spinless fermions). 
Even in this simpler case, the interplay between Coulomb repulsion 
and the kinetic energy is a complicated issue. When many electrons are 
inside the quantum volume $a_B^2$, $r_s$ is small and the 2DEG is a 
Fermi liquid, with a Fermi energy much larger than the Coulomb 
energy. When $r_s$ is large, the volume per electron $a^2$ is 
large compared to $a_B^2$, and one has an electron solid (Wigner crystal) 
with weak quantum effects. Between those two limits, the nature of the 
ground state remains unclear. 

From Quantum Monte Carlo studies, it is generally believed that there 
is a single first order liquid-solid transition. However this result 
is not free of certain assumptions, because of the well known ``sign 
problem'' of the Monte Carlo methods applied to fermions. One reason 
to question the existence of a single transition comes from general 
considerations about Landau theory of phase transitions and estimates 
of interface energies. According to Ref. \cite{spivak2}, a single first 
order transition is ruled out because a macroscopic phase separation 
between a liquid and a solid is unstable. A way to fix the sign problem 
of the Monte Carlo approaches consists in imposing the GS nodal structure. 
Using a fixed node approach, two nodal structures (a Slater determinant 
of plane waves for the liquid, of localized orbitals for the solid) have 
been compared \cite{ceperley}, giving $r_s^W \approx 37$ for the critical 
density at which the transition occurs between the two assumed nodal 
structures. However, one cannot exclude the existence of better nodal 
structures giving lower GS energies when $r_s$ is neither small nor large. 
And indeed, a third nodal structure has been recently studied \cite{fw,f} 
in a fixed node Monte Carlo approach, using Bloch waves in a variational 
attractive potential located at the sites of the Wigner lattice instead 
of free plane waves. This third nodal structure gives a lower GS energy 
for $30 < r_s < 80$ and  $N \rightarrow \infty$, supporting the existence 
of a new hybrid liquid-solid phase. Since one does not know the nodal 
structures, the Monte Carlo methods make problem, and exact numerical 
studies of small systems remain useful, despite the presence of large 
finite size effects. They could provide a better understanding of the 
low energy physics of a 2DEG, suggesting better trial wave functions for 
fixed node Monte Carlo methods. Or they could be the starting point of a 
finite size scaling theory \cite{bwp3} for the interacting 2DEG. This is 
the road which we have explored in a series of works 
\cite{bwp1,bwp2,sp,ksp,np,mp,fnp} and which we continue here.
  
 To do an exact study, we take $N$ electrons on a $L \times L$ lattice 
for having an Hilbert space of finite size. If $N$ and $L$ are small 
enough, the GS can be obtained by Lanczos algorithm. Beside the finite 
size effects due to the finite value of $N$, one has the lattice effects 
due to the finite value of $L$. If the lattice effects are irrelevant, 
the lattice GS behaves as the continuum GS. Notably, the parameter $r_s$ 
remains a scaling parameter. If the lattice effects are relevant, $r_s$ 
becomes meaningless, the scaling behavior of the continuum limit ceases 
to exist for giving rise to a new lattice physics. In the absence of 
disorder, the lattice effects were studied \cite{fnp} as a function of 
the strength $U$ of the Coulomb repulsion. The strength $U^*$ above which 
the lattice GS and the continuum GS become different was given. The 
previous studies of a few particle systems \cite{ksp,np,fnp} give the 
following picture: If one increases $U$ for a sufficiently small number 
$N$, the Fermi system disappears above $U^F$ to become a Wigner solid 
above $U^W$, through an intermediate regime for $U^F< U <U^W$ which 
could be the microscopic trace of an ``electron supersolid'', before 
exhibiting strong lattice effects above $U^*$. The Wigner solid without 
lattice effects can be described by a continuum expansion where the small 
parameter is $1/\sqrt{r_s}$. For a lattice model, 
$r_s=(UL/t)(1/\sqrt{4 \pi N})$, where the hopping term $t$ sets the scale 
of the lattice bandwidth without interaction. Above $U^*$, the continuum 
expansion breaks down and the Wigner solid with lattice effects is described 
by a different expansion in power of another small parameter $t/U$. If 
$N$ is large, the lattice effects can appear in the Fermi regime, a case 
which we will not consider here. In Ref. \cite{fnp}, the different regimes 
have been illustrated by numerical studies using $N=3$ spinless fermions 
in lattices of size $L=6$ up to $18$. Three criteria where given to get 
$U^*$ and check for $N=3$. The numerical results have been reproduced by 
a continuum $1/\sqrt{r_s}$ expansion of the Wigner solid for $U<U^*$ and 
by a lattice $t/U$-expansion for $U>U^*$. Since it is possible to extend 
these expansions to an arbitrary number $N$ of particles, the lattice 
threshold $r_s^*$ below which the lattice effects are relevant was given 
in the thermodynamic limit as a function of $s/a_B$, $s$ being the lattice 
spacing. The purpose of this work is to use the persistent current as a 
tool for characterizing the effect of a weak disorder upon the four 
regimes which we have identified without disorder. The considered system 
is a $2d$ torus pierced by an Aharonov-Bohm flux. The random potentials 
remove the various translational, rotational, and inversion symmetries 
exhibited by the non-disordered lattice. The persistent currents display 
new regimes induced by the interplay between electronic correlations and 
random potentials and show more precisely the nature of the supersolid GS. 
In the numerical part of this work, we take $N=3$ spinless fermions for 
lattices of sizes $L=6$ and $9$. When possible, our numerical results 
are reproduced by analytical expansions which can be generalized to more 
particles. 

\subsection{Summary of the main results}
\label{subsection1.2}

 This paper is organized as follows: the disordered lattice 
model is introduced in section \ref{section2}. The persistent 
currents (local, total, transverse, longitudinal) yielded by 
an enclosed Aharonov-Bohm flux $\Phi$ are defined in section 
\ref{section3}. A relation between the expectation value 
$\left<{\bf K}\right>$ of the total momentum and the current $I$ 
is derived in section \ref{section4} for the continuum limit of a 
lattice of arbitrary dimension $d$. Since this  relation is only 
valid when the one particle states of high momenta ${\bf k}$ are 
empty, the continuum and the lattice regimes can be defined from 
the study of $I$. In the continuum regime, $I$ is invariant when 
${\bf K}$ is invariant. In the lattice regime, $I$ can decay while 
${\bf K}$ remains invariant. In the continuum regime, the lattice GS 
behaves as in the continuum limit and displays the same universal scaling 
laws. In the lattice regime, the lattice GS becomes different and there 
is no scaling. 

Hereafter, we assume $N=3$ spinless fermions and we vary the lattice 
parameters $t$ (kinetic energy), $U$ (Coulomb repulsion), $W$ (disorder) 
and $L$ (lattice size). In section \ref{section5}, we review the 4 
regimes obtained when $U/t$ increases without disorder ($W/t=0$) 
for a sufficient value of $L$: continuum Fermi system ($U<U^F$), 
continuum supersolid ($U^F<U<U^W$), continuum Wigner molecule 
($U^W<U<U^*$) and lattice Wigner molecule ($U^*<U$). 
The effects of a very weak disorder are described in section 
\ref{section6}. One gets 3 lattice regimes as $U$ exceeds $U^*$. 
First the disorder remains negligible and one gets a Ballistic 
Wigner Molecule (BWM regime). When $U$ reaches a value $U_{stripe}>U^*$, 
a new lattice regime characterized by a Coulomb Guided Stripe of Current 
(CGSC) is found when $U_{stripe}<U<U_{loc}$. Above $U_{loc}$, one gets 
a Localized Wigner Molecule (LWM regime). In the CGSC regime, $I$ flows 
along the axis of the Wigner molecule instead of flowing along the shortest 
direction enclosing the flux, as when $U<U_{stripe}$ (BWM regime) or 
when $U>U_{loc}$ (LWM regime). The perturbation theories describing these 
three lattice regimes are given, and their range of validity are estimated. 
In section \ref{section7}, figures where one can see how $I$ depends on 
$U$, $W$ and $t$ are shown, exhibiting the decay of $I$ predicted by the 
three lattice perturbation theories, and the continuum regimes where $I$ is 
invariant when ${\bf K}$ is invariant. In section \ref{section8}, the 
effect of the disorder upon the continuum-lattice crossover is studied. 
We also show how an avoided level crossing induces a jump of $I$ and a 
change of its sign when $N=3$, in a continuum regime where $I$ is 
otherwise invariant. The phase diagram of the different lattice regimes 
obtained using $N=3$ spinless fermions is sketched in the plane 
$\left(U/t,W/t\right)$ (Fig \ref{Fig10}). In section \ref{section9}, we 
study the effect of random potentials in continuum regimes where $I$ 
remains invariant. We give a detailed study of the case $W=t=1$, where 
the motion is diffusive without interaction for the considered values of 
$L$. The study of the map of local currents shows that the stripe of 
current observed in the lattice CGSC-regime persists in the continuum 
limit. The persistence of 1d motions yielded by the Coulomb repulsion 
in a 2d disordered lattice allows us to obtain the parity of the number 
of particles contributing to $I$ from the sign of $I$. This leads us to 
suggest that a $N=3$ diamagnetic Wigner molecule melts in a disordered 
lattice through an intermediate ``supersolid'' regime where a $N=2$ 
paramagnetic Wigner molecule co-exists with a third nearly localized 
particle, before becoming a Fermi glass. In section \ref{section9}, 
the three first harmonics of the function $I(\Phi)$ are given as a 
function of $r_s$, showing in greater details how a 2d current $I$ of 
random sign becomes a 1d current of given sign when the electronic 
correlations increase. In section \ref{section10}, we underline the main 
results obtained in studying three particles: the existence of a regime 
of stripe and of a supersolid regime where the zero point motion of the 
electron solid becomes of the order of the system size without disorder, 
or where a delocalized pair co-exists with a localized particle with 
disorder. We conclude by discussing the possible relevance of our results 
when $N$ becomes larger.

\section{Lattice Hamiltonian with disorder}
\label{section2}

 The lattice Hamiltonian $H_l$ describing $N$ polarized electrons 
free to move on a $L \times L$ disordered square lattice with 
periodic boundary conditions (BCs) reads:    
\begin{equation}
H_l=H_{kin}+H_{int}+H_{dis}. 
\label{H-lattice-site}
\end{equation}
We write these three terms using the creation (annihilation) 
operators in real space and in momentum space.

\subsection{Site basis}
\label{subsection2.1}

Using the operators $c_{\bf{j}}^{\dagger}$ ($c_{\bf{j}}$) which 
create (annihilate) a polarized electron (spinless fermion) at the 
lattice site $\bf{j}$, the kinetic term reads:
\begin{equation}
H_{kin}=t \left(4N- \sum_{\left<{\bf j},{\bf j'}\right>}
c_{\bf j}^{\dagger} c_{\bf j'}\right),  
\label{H-lattice-kinetic}
\end{equation}
the Coulomb term reads:
\begin{equation}
H_{int}=\frac{U}{2}  \sum_{{\bf j} \neq {\bf j'}} 
\frac{n_{\bf j} n_{\bf j'}}
{|d_{\bf jj'}|}.  
\label{H-lattice-interaction}
\end{equation}
The third term describes a random substrate: 
\begin{equation}
H_{dis} = W \sum_{\bf j} \epsilon_{\bf j} n_{\bf j}.  
\label{H-lattice-disorder}
\end{equation}
$W$ gives the strength of the disorder and the variables 
$\epsilon_{\bf j}$ are taken at random between $[-1/2,1/2]$. 
$\left< {\bf j},{\bf j'}\right>$ means that the sum is restricted to 
nearest neighbors. $n_{\bf j}=c_{\bf{j}}^{\dagger}c_{\bf{j}}$. 
For a square of area $D^2$, the lattice size is given by $L=D/s$, 
where $s$ is the lattice spacing. Assuming periodic BCs, we define 
the distance $d_{\bf jj'}$ between the sites ${\bf j}$ and ${\bf j'}$ 
in unit of $s$, as  
\begin{equation}
d_{\bf jj'}=\frac{L}{\pi} \sqrt{\sin^2\frac{|d_{x}|\pi}{L}+
\sin^2\frac{|d_{y}|\pi}{L}}.
\label{distance2}
\end{equation}
This metric coincides at short distance with the natural $2d$ metric 
and avoids \cite{fnp} to have singular repulsions with cusps when 
$d_{\bf jj'}\approx L/2$ and $s \rightarrow 0$. The corresponding 
Coulomb repulsion is essentially equivalent to the one obtained 
from Ewald's summation, where one assumes the infinite periodic 
repetition of the same $L \times L$ square instead of a $2d$ torus.

\subsection{Plane wave basis}
\label{subsection2.2}

 The Hamiltonian (\ref{H-lattice-site}) can also be written using the 
operators $d_{\vec{k}}^{\dagger}$ ($d_{\vec{k}}$) creating (annihilating) 
a polarized electron in a plane wave state of momentum $\vec{k}$: 
\begin{equation}
 d_{\bf k} = \frac{1}{L} \sum_{\bf j} e^{-i {\bf k j}} c_{\bf j}. 
\end{equation}
\begin{equation}
H_{kin}=4Nt - 2t \sum_k \left(\cos k_x +\cos k_y \right)d_{\bf k}^\dagger
d_{\bf k}  
\label{Hamiltonian-kin-k}
\end{equation}
\begin{equation}
H_{int}= U \sum_{{\bf k},{\bf k'},{\bf q}} V({\bf q}) 
d_{{\bf k}+{\bf q}}^\dagger 
d_{{\bf k'}-{\bf q}}^\dagger d_{\bf k'} d_{\bf k}
\label{Hamiltonian-int-k}
\end{equation}
\begin{equation}
H_{dis}=W \sum_{{\bf k},{\bf q}} \epsilon({\bf q}) 
d_{{\bf k}+{\bf q}}^\dagger d_{\bf k}
\label{Hamiltonian-dis-k}
\end{equation}
where
\begin{equation}
V({\bf q})={1\over 2 L^2} \sum_{\bf j} {\cos{\bf qj}\over d_{\bf j0}}. 
\end{equation}
and
\begin{equation}
\epsilon({\bf q})={1\over L^2} \sum_{\bf j} \epsilon_{\bf j} 
\exp(-i{\bf q \cdot j}). 
\end{equation}

\begin{figure}[ht]  
\vskip.2in
\centerline{\epsfxsize=8cm\epsffile{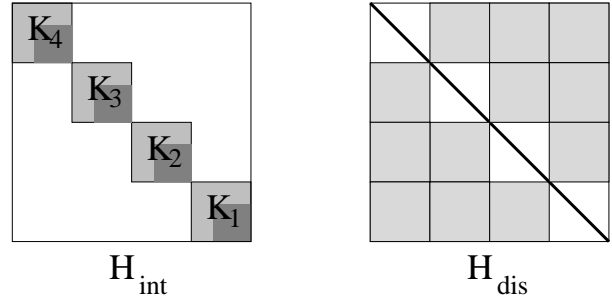}} 
\caption{Structures of the interaction and disorder Hamiltonian 
matrices written in the basis of the PWSDs. The kinetic energy 
gives a diagonal matrix. We have sketched the structures 
for 4 subspaces of total momenta ${\bf K}_1, \ldots, {\bf K}_4$. 
The matrix elements are zero in the white parts. For $H_{int}$, 
the darker regions corresponds to hopping terms between PWSDs 
built out from one body states of small momenta $\bf k$ only. 
In the shaded parts of $H_{int}$ and $H_{W}$, the matrices are 
sparse.} 
\label{Fig3}
\end{figure} 

In Fig. \ref{Fig3}, the structure of the Hamiltonian matrix 
for $N$ spinless fermions is sketched in the basis of the Plane Wave 
Slater Determinants (PWSDs) $d_{\bf k_1}^\dagger \ldots 
d_{\bf k_N}^{\dagger} \left|0\right>$. The matrix corresponding to 
$H_{kin}$ is diagonal. If we order the PWSDs by series of same total 
momentum ${\bf K}=\sum_{i=1}^N{\bf k_i}$, the matrix corresponding to 
$H_{int}$ is block-diagonal, each block being characterized by the 
same momentum ${\bf K}$. Inside each block, we order the PWSDs by 
increasing kinetic energy. The matrix elements between PWSDs for 
which all the components of ${\bf k}$ are small such that $\sin k_x 
\approx k_x$ and $\sin k_y \approx k_y$, are indicated by a darker grey. 
Moreover, only PWSDs of same ${\bf K}$ having $N-2$ momenta ${\bf k}$ in 
common out of $N$ can be directly coupled by the two-body interaction. When 
$N\geq 3$, this means that each block of given ${\bf K}$ is sparse. In 
contrast, the random matrix due to $H_{W}$ has zero matrix elements inside 
the diagonal blocks of momenta ${\bf K}$, but gives non zero off-diagonal 
terms coupling those diagonal blocks. These coupling off-diagonal blocks 
are even sparser, having non zero terms between PWSDs differing by a 
single ${\bf k}$ only, out of $N$. 

\subsection{Lattice factor $r_s$}
\label{subsection2.3}

The continuum thermodynamic limit is obtained when the finite size effects 
due to the finite value of $N$ and the lattice effects due to the finite 
value of $L$ are negligible. This is clearly out of reach of the Lanczos 
algorithm. However, for a finite value of $N$, $r_s$ remains a scaling 
parameter when the lattice effects are negligible. In a medium of dielectric 
constant $\varepsilon$, the strength $U$ of the Coulomb repulsion reads:
\begin{equation}
U= \frac{e^2}{\varepsilon s} 
\end{equation}
while 
\begin{equation}
t= \frac{\hbar^2}{2m^* s^2} 
\end{equation}
is the hopping element between nearest neighbor sites and sets 
the bandwidth of carriers of effective mass $m^*$. Taking $m=m^*$ 
and $\varepsilon=1$, one has in lattice units $1 Ry=U^2/4t$, 
$a_B=2st/U$ and 
\begin{equation}
r_s=\frac{a}{a_B}=\frac{UL}{2t\sqrt{\pi N}} 
\label{r_s-lattice}
\end{equation}  
for the rydberg, the Bohr radius and the factor $r_s$ respectively.

Instead of changing the number $N$ of particles, one can vary $r_s$ 
keeping $N$ fixed and varying $U$, $t$ and $L$, and hence the 
parameter 
\begin{equation}
r_l=\frac{UL}{t}= r_s \sqrt{4 \pi N}.
\label{r_l-lattice}
\end{equation}  
This amounts to change $r_s$ by changing $a_B$ instead of $a$. 
The lattice effects are important \cite{fnp} either at large 
values of $U$ (hence large $r_l$) for a given $N$, or at large 
values of $N$ (hence small $r_s$) for a given $U$. Hereafter, we 
use $r_l$ or $U$ instead of $r_s$ in order to avoid confusion.

\section{Persistent current $I$}
\label{section3}

\begin{figure}[ht]  
\vskip.2in
\centerline{\epsfxsize=8cm\epsffile{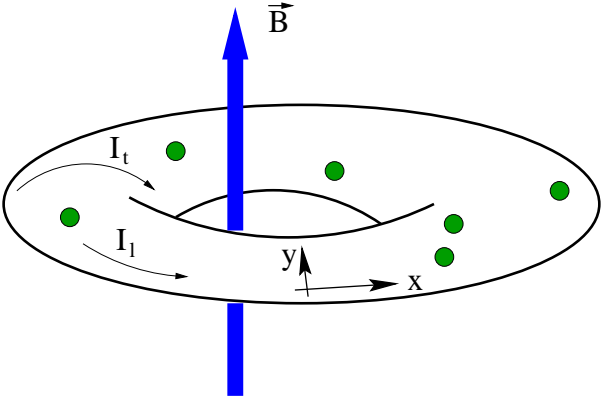}} 
\caption{2d Torus with $N$ electrons enclosing an Aharonov-Bohm flux 
$\Phi \propto B$} 
\label{fig-m2}
\end{figure}

 An Ahronov Bohm flux $\Phi$ is enclosed around the longitudinal 
$x$-direction as sketched in Fig. \ref{fig-m2}. 

By a gauge transformation, $\Phi$ can be included in the longitudinal BC 
(antiperiodic BC corresponding to $\Phi=\Phi_0/2$ in our convention), 
while the BC remains always periodic in the transverse $y$-direction. 
This flux creates persistent currents for a GS of wave-function 
$\left|\Psi_0\right>$ and of energy $E_0(\Phi)$:  

\begin{itemize}

\item At a site ${\bf j}=(j_x,j_y)$, the local longitudinal current 
$j^{l}_{\bf j}(\Phi)$ and the local transverse current 
$j^{t}_{\bf j}(\Phi)$ read
\begin{equation}
j^{l}_{\bf j}(\Phi)= 2 \Im \left<\Psi_0\left|c_{\bf j}^\dagger 
c_{(j_x+1,j_y)} e^{2 \pi i {\Phi\over L \Phi_0}} \right|\Psi_0\right>, 
\end{equation}
\begin{equation}
j^{t}_{\bf j}(\Phi)= 2 \Im \left<\Psi_0\left|c_{\bf j}^\dagger 
c_{(j_x,j_y+1)} \right|\Psi_0\right> 
\end{equation}
in units of $\sqrt{t/2m}/s$. 

\item The total longitudinal current $I_{l}(\Phi)$ enclosing the 
flux is obtained by summing the local longitudinal currents: 
\begin{equation}
I_{l}(\Phi) = -\frac{\partial E_0 (\Phi)} {\partial\Phi} 
            =   \sum_{j_y=1}^L j^{l}_{\bf j} 
             = {1\over L} \sum_{\bf j} j^{l}_{\bf j}.
\end{equation}
The latter equation comes from the conservation of the current. 
The total transverse current $I_{t}(\Phi)$ is given by a similar 
formula using the $j^{t}_{\bf j}\Phi)$. 

\item The change $\Delta E_0(\Phi)$ of the GS energy $E_0$ 
induced by $\Phi$: 
\begin{equation}
\Delta E_0(\Phi)=E_0(0)-E_0(\Phi)
\label{E-change}
\end{equation}
is a quantity related to $I$ which is simpler to study. 

\end{itemize}

 The GS total longitudinal current
\begin{equation}
I_{l}(\Phi)= \frac{2}{L} \Im \sum_{\bf j}\left<\Psi_0\left|c_{\bf
j}^\dagger c_{(j_x+1,j_y)} e^{2 \pi i {\Phi\over L \Phi_0}}
\right|\Psi_0\right>
\end{equation}
can be expressed in terms of the operators $d_{\vec{k}}^{\dagger}$ 
($d_{\vec{k}}$). Hereafter, the $k_x$ and $k_y$ components are quantized 
in multiples of $2\pi/L$, and an explicit phase factor 
$\exp(2\pi \Phi/ (\Phi_0 L))$ is added to the $k_x$-components. 
This gives 
\begin{equation}
I_{l}(\Phi) = \frac{2}{L} \Im \sum_{\bf k} 
e^{i \left(k_x+ 2 \pi {\Phi \over L \Phi_0} \right)} 
\left<\Psi_0\left|d^{\dagger}_{\bf k} d_{\bf k}\right| \Psi_0 \right>
\label{current-reciprok}
\end{equation}
for the total longitudinal current and 
\begin{equation}
I_{t}(\Phi) = \frac{2}{L} \Im \sum_{\bf k} 
e^{i k_y} 
\left<\Psi_0\left|d^{\dagger}_{\bf k} d_{\bf k}\right| \Psi_0 \right>
\end{equation}
for the total transverse current. 

 Expressed in terms of its projections $\Psi_{\bf k_1,\ldots,k_N}$ 
onto the PWSDs $\prod_{n=1}^{N}d_{\bf k_n}^\dagger \left|0\right>$, 
the GS reads: 
\begin{equation}
\left|\Psi_0\right>=\sum_{\bf k_1,\ldots,k_N} 
\Psi_{\bf k_1,\ldots,k_N} \prod_{n=1}^{N}d_{\bf k_n}^\dagger 
\left|0\right>
\end{equation}
which gives
\begin{eqnarray}
I_{l}(\Phi) &=& \frac{2}{L} \sum_{\bf k_1,\ldots,k_N} \left| 
\Psi_{\bf k_1 \ldots k_N} \right|^2 
\sum_{i=1}^N \sin{ \left(k_{ix}+ 2 \pi {\Phi\over L \Phi_0}
\right)}  \nonumber \\
I_{t}(\Phi) &=& \frac{2}{L} \sum_{\bf k_1,\ldots,k_N} \left| 
\Psi_{\bf k_1 \ldots k_N} \right|^2 
\sum_{i=1}^N \sin{k_{iy}}.
\label{sinus-expr-lon-trans}
\end{eqnarray}

\section{Relation between $I$ and ${\bf K}$ in the continuum 
limit and lattice threshold $U^*$}
\label{section4}
 
 As far as the lattice GS occupies small momenta ${\bf k}$, 
one can approximate the dispersion relation of the lattice model 
by the parabolic dispersion relation of the continuum limit: 
\begin{equation}
2-2 \cos k_i \approx k_i^2  \ \ \ and \ \ \ \sin k_i \approx k_i.  
\label{cont-approx}
\end{equation}
In this approximation, the expressions (\ref{sinus-expr-lon-trans}) 
become
\begin{eqnarray}
I_{l}(\Phi) &\approx& \frac{2}{L} \sum_{\bf k_1,\ldots,k_N}
\left| \Psi_{\bf k_1,\ldots,k_N} \right|^2 \sum_{i=1}^N \left(  
k_{ix}+ {2 \pi \Phi\over L \Phi_0} \right) \\
I_{t}(\Phi) &\approx& \frac{2}{L} \sum_{\bf k_1,\ldots,k_N}
\left| \Psi_{\bf k_1,\ldots,k_N} \right|^2 \sum_{i=1}^N k_{iy}  
\end{eqnarray}
which give
\begin{eqnarray}
I_{l}&=& \frac{2}{L} \left(\left<K_x\right> + {2 \pi N \over L}{\Phi\over
\Phi_0}\right) \\
I_{t}&=& \frac{2}{L} \left<K_y\right>,
\label{lat-inv}
\end{eqnarray}
where $\left<K_{x,y}\right>= \sum_{\bf k_1,\ldots,k_N} 
\left( \sum_{i=1}^N k_{i,x,y} \right) \left| 
\Psi_{\bf k_1,\ldots,k_N}\right|^2$.

 This proves an important theorem which was known in one dimension 
and numerically checked \cite{muller-groeling,krive,burmeister} in 
two dimensions: the GS persistent current $I$ is independent of the 
interaction strength $U$ in the continuum limit of a non-disordered 
lattice model of arbitrary dimensions. Moreover, this provides one 
criterion (criterion 2 of Ref. \cite{fnp}) for obtaining the lattice 
threshold $U^*$. We have shown that $I$ is proportional to ${\bf K}$ 
when the momenta ${\bf k}$ occupied by the GS are small, allowing to 
use the approximation (\ref{cont-approx}). The invariance of $I$ is a 
direct consequence of the invariance of ${\bf K}$. When $U=W=0$, the 
approximations (\ref{cont-approx}) require small filling factors $N/L^2$. 
If one turns on $U$ for $W=0$ and a small filling factor, 
$I \propto {\bf K}$ remains independent of $U$ when $U$ is smaller than 
the threshold $U^*$. When $U>U^*$, the GS of a non-disordered lattice 
begins to occupy, inside the same subspace of total momentum ${\bf K}$, 
large one particle momenta ${\bf k}$, and the relation between $I$ and 
${\bf K}$ ceases to be valid. Without disorder, ${\bf K}$ is always 
conserved, and $I$ remains independent of $U$ as far as 
$I \propto {\bf K} $. The breakdown of the invariance of $I$ is a 
consequence of a localization-delocalization crossover induced by 
$U$ in momentum space. For a fixed value of $t$, this crossover can be 
induced by $U$ inside a subspace of given total momentum ${\bf K}$, 
by $W$ among subspaces of different ${\bf K}$, or by the combined role 
of $U$ and $W$. The regime where ${\bf K}$ is conserved defines the 
ballistic many body regime. 
 
\section{Overview of the non disordered case for $N=3$}
\label{section5}

Before considering the case where random potentials are included, 
it is useful to summarize what we know when there is no disorder. 
Let us review a few published \cite{np,fnp} and unpublished results 
based on a study of the case $N=3$.

\subsection{Continuum Wigner molecule above $U^W$ \\ 
($r_s > r_s^W \approx 50$)}
\label{subsection5.1}

 On a continuum $D \times D$ square with periodic BCs, the Coulomb 
repulsion with the taken $2d$ metric reads 
\begin{equation}
V({\bf r})= {e^2 \pi\over D 
\sqrt{\sin^2 {r_x \pi\over D}+
\sin^2 {r_y \pi \over D}}}.
\label{sinus-potential}
\end{equation}
The GS center of mass is delocalized. When the density $N/D^2$ is small 
enough, the spacings $d$ between the particles are large and one obtains 
a Wigner molecule, which is in the continuum limit if the fluctuations 
of $d$ are also larger than $s$. For a given center of mass, 
to put the particle on the sites ${\bf r}_1=(0,0)$,  ${\bf r}_2=(D/3,D/3)$ 
and ${\bf r}_3=(-D/3,-D/3)$ yields for the Coulomb energy a minimum value:  
\begin{equation}
E_{Coul} = \frac{\sqrt{6} e^2 \pi}{D}.
\end{equation}
For a large $D$, one can expand the pair-potential around the equilibrium 
inter-particle spacing ${\bf r}_0=(D/3,D/3)$ up to the second order to 
get harmonic oscillations around the equilibrium positions. The 
Hamiltonian $H_c \approx E_{Coul}+H_{harm}$, where the harmonic 
part reads:
\begin{equation}
H_{harm} = {1\over 2 m} \sum_{i=1}^3 {\bf p}_i^2 
+ {\vec X} {\hat M} {\vec X}.
\label{Hamilton-expansion}
\end{equation}
The vector ${\vec X}=(x_1,y_1,x_2,y_2,x_3,y_3)$ describes 
the motions of the molecule around equilibrium. The $6\times 6$ 
matrix ${\hat M}$ is given by:
\begin{equation}
{\hat M} = \left(\matrix{2A&2B&-A&-B&-A&-B\cr
2B&2A&-B&-A&-B&-A\cr -A&-B&2A&2B&-A&-B\cr 
-B&-A&2B&2A&-B&-A\cr -A&-B&-A&-B&2A&2B\cr
-B&-A&-B&-A&2B&2A}\right).
\end{equation}
where 
\begin{equation}
A = {7\sqrt{6}\over 72}{e^2 \pi\over D^3} 
\end{equation}
and $B = 3A/7$. 
Diagonalizing ${\hat M}$, one obtains the normal modes, 
while the eigenvalues of ${\hat M}$ give their frequencies. 

One obtains two eigenvectors ${\vec \chi_1}$ and ${\vec \chi_2}$  
of eigenvalue $0$, which corresponds to the translation of the 
center of mass of the molecule, two other eigenvectors 
${\vec \chi_3}$ and ${\vec \chi_4}$ of eigenvalue $10B$, 
which corresponds to the longitudinal mode, while the two last 
eigenvectors ${\vec \chi_5}$ and ${\vec \chi_6}$ of eigenvalue $4B$ 
correspond to the transverse mode. The expressions of eigenvectors 
${\vec \chi_i}$ are given in Ref. \cite{fnp}.

Using these normal coordinates, the Hamiltonian (\ref{Hamilton-expansion}) 
becomes a decoupled sum of four harmonic oscillators:
\begin{equation}
H_{harm}-{\hbar^2 \over 2 m}\sum_{\alpha=1}^6 {\partial^2\over 
\partial\chi_\alpha^2} = 10B (\chi_3^2+\chi_4^2) + 4B 
(\chi_5^2+\chi_6^2), 
\end{equation}

Without disorder, the interaction can only couple states of 
same total momentum ${\bf K}$. Inside a subspace of a total 
momentum ${\bf K}$, the wave functions can be factorized as 
\begin{equation}
\Psi(\chi_1,\dots,\chi_6) =\Psi^{\bf K}_{cm} (\chi_1,\chi_2) 
\Psi^{n_3,\dots,n_6}_{rel} (\chi_3,\dots,\chi_6). 
\end{equation}
The kinetic energy associated to the motion of the center mass 
reads: 
\begin{equation}
E_{cm}({\bf K}) = -{\hbar^2 \over 2 N m} {\bf K}^2.  
\end{equation}
The wave-function $\Psi^{n_3=0,\dots,n_6=0}(\chi_3,\dots,\chi_6)$
corresponding to the relative motions can be factorized as:
\begin{equation}
\Psi^{n_3=0,\dots,n_6=0}_{rel} = 
\varphi_{0L}(\chi_3) \varphi_{0L}(\chi_4)\varphi_{0T}(\chi_5) 
\varphi_{0T}(\chi_6)
\end{equation}
for the ground state ($n_3=0,\dots,n_6=0$) of momentum ${\bf K}$.
$L,T$ refer to the transverse and longitudinal modes and $\varphi_0$ 
is the GS of an harmonic oscillator:
\begin{equation}
\varphi_0(x)= {1\over l_\omega^{1/2} \pi^{1/4}} 
\exp-\frac{x^2}{ 2l_\omega^2}, 
\end{equation}
of characteristic length: 
\begin{equation}
l_\omega = \left(\frac{\hbar^2}{m^2 \omega^2}\right)^{1/4}. 
\end{equation}
For the expanded pair potentials, the GS energy $E_0({\bf K})$ 
inside a subspace of momentum ${\bf K}$ becomes 
\begin{equation}
E_0({\bf K})= E_{Coul} + E_{cm}({\bf K}) + \hbar (\omega_T+\omega_L); 
\end{equation}
where 
\begin{equation}
\omega_L = \sqrt{\frac{20B}{m}} \ \ , \ \omega_T = \sqrt{\frac{8B}{m}}.
\end{equation}
In lattice units, one gets:  
\begin{equation}
E_0({\bf K}) = E_{Coul} + E_{cm}({\bf K}) + E_{rel,0}, 
\label{harm-theo}
\end{equation}
where $E_{rel}(n_3=0,\dots,n_6=0)=E_{rel,0}$. 
\begin{eqnarray}
E_{cm}({\bf K}) &=& \frac{t}{N} {\bf K}^2 \\
E_{rel,0} &=& \frac{\sqrt{6} \pi U}{L}
+ (\sqrt{5}+\sqrt{2}) \sqrt{\sqrt{6}\pi^3\over 3} \sqrt{Ut \over L^3}.
\end{eqnarray}

$E_0({\bf K})$ depends on ${\bf K}$ only through $E_{cm}({\bf K})$, 
while $E_{Coul}$ and the transverse and longitudinal modes remain 
independent of ${\bf K}$. This is still the case if one goes 
beyond the harmonic approximation for the pair potentials, including 
anharmonic corrections. The only approximation which we have done 
is to assume that the relative motions are small compared to $D$. 
Thus the GS relative wave functions $\Psi^{n_3=0,\dots,n_6=0}_{rel} 
(\chi_3,\dots,\chi_6)$ and $E_{rel}(n_3=0,\dots,n_6=0)$ are approximated 
by the solutions of an unbounded $2d$ system, while the wave functions of 
the center of mass $\Psi^{\bf K}_{cm} (\chi_1,\chi_2)$ and $E_{cm}({\bf K})$ 
correspond to a bounded $2d$ system. The assumed $2d$ harmonic oscillations 
make sense only if the Wigner molecule is rigid enough for making  
negligible the boundary effects upon the relative motions. 
 
\subsection{Supersolid molecule below $U^W$ \\
($r_s^F< r_s < r_s^W$)}
\label{subsection5.2}

\begin{figure}[ht]  
\vskip.2in
\centerline{\epsfxsize=8cm\epsffile{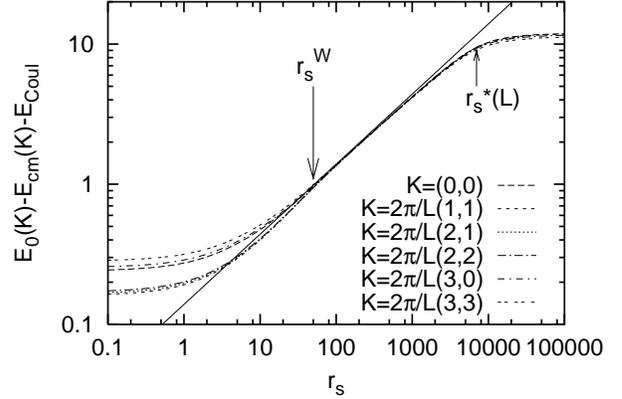}} 
\caption{$W=0$, $L=18$ and $N=3$: The GS-energies 
$E_0({\bf K})-E_{cm}({\bf K})-E_{Coul}$ for different  
total momenta ${\bf K}$ as a function of $r_s$. The arrow shows 
$r_s^W \approx 50$ above which the behavior is given by 
Eq \ref{harm-theo} (solid line $0.139 \sqrt{r_s}$) up to 
the lattice threshold $r_s^*$.
} 
\label{Fig1}
\end{figure}

 To determine the values of $r_s$ where the GS energy is described by 
Eq (\ref{harm-theo}), we plot in Fig. \ref{Fig1} the GS energies 
$E_0({\bf K})-E_{cm}({\bf K})-E_{Coul}$ as a function of $r_s$, 
for a size $L=18$ and different subspaces of momentum ${\bf K}$. 
$E_{Coul}=\sqrt{6} \pi U/L$ when $L/3$ is integer.

The solid line ($0.139 \sqrt{r_s}$) gives the behavior implied 
by the continuum harmonic expansion (Eq \ref{harm-theo}) with unbounded 
relative motions. One can see that this theory is accurate for 
$r_s^W < r_s < r_s^*$. Above $r_s^W \approx 50$, the relative motions 
become localized and coincide to unbounded harmonic vibrations. 
Below $r_s^W$, the behavior is more complicated. On one side, the 
$E_0({\bf K})$ are roughly described above $r_s^F \approx 10$ by the 
expansion (\ref{harm-theo}) valid for a continuum solid molecule. On the 
other side, the relative motions depend on ${\bf K}$, a dependence 
which cannot exist unless the relative motions are extended over a scale 
of the order of the system size $D=Ls$. Large zero-point motions are 
expected in a supersolid. For $N=3$, we have shown that there is regime 
where there is a floppy Wigner molecule, though the relative motions 
depends on the quantization of ${\bf K}$ and hence on the BCs. This 
intermediate ``supersolid'' regime was noted in Ref. \cite{np} using 
another $2d$ metric for $L=6$, a limit where the lattice effects play 
a role. Our results for a Wigner molecule describable by a continuum 
theory when $L=18$ gives further support to the existence of an 
intermediate regime for $r_s^F <r_s < r_s^W$ which is not a lattice 
effect. Hereafter, we use the word ``supersolid'' to refer to this regime.  

\subsection{Lattice regime above $U^*(L)$ \\
($r_l > r_l^*(L)$)}
\label{subsection5.3} 
 
\begin{figure}[ht]  
\vskip.2in
\centerline{\epsfxsize=8cm\epsffile{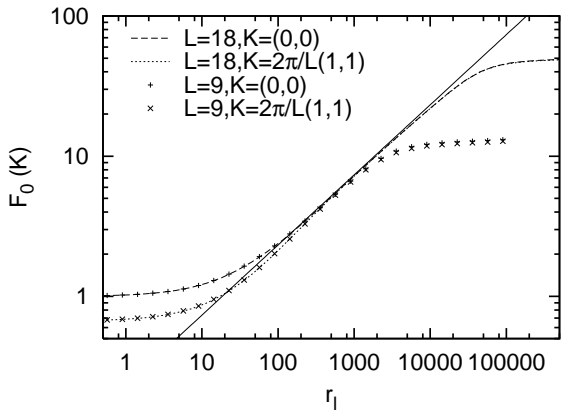}} 
\caption{$W=0$, $N=3$: The dimensionless GS-energies 
$F_0$ for different total momenta ${\bf K}$ as a 
function of $r_l=UL/t$ for $L=9$ and $L=18$. The solid 
line $0.2326 \sqrt{r_l}$ comes from Eq (\ref{harm-theo-F0}).} 
\label{Fig1.a}
\end{figure}

 The GS energy of the lattice model (Hamiltonian \ref{H-lattice-site}) 
is also not described by the continuum expansion (\ref{harm-theo})  
above the lattice threshold $r_l^*(L)$. Three equivalent criteria were 
introduced in Ref. \cite{fnp} for giving $r_l^*(L)$. One was based on 
the breakdown of the invariance of $I$ discussed in section \ref{section4}. 
Let us introduce a fourth criterion which gives a similar answer: the 
lattice spacing $s$ becomes relevant when the size $l_{\omega}$ of the 
harmonic oscillations becomes of the order of the lattice spacing $s$. 
The longitudinal modes giving the smallest scale, the criterion reads  
\begin{equation}
l_{\omega_L}= \left(\frac{\hbar^2}{m^2 {\omega}_L^2}\right)^{1/4} 
\approx s, 
\label{criterion4}
\end{equation}
giving $r_l^* (L) \approx (2\sqrt{6})/(5\pi) L^4$ or 
\begin{equation}
\frac{U^*(L)}{t} \approx \frac{2\sqrt{6}}{5\pi} L^3.
\label{criterion4-bis}
\end{equation}
When $U>U^*(L)$ ($r_l>r_l^*(L)$), the lattice model exhibits a 
different behavior than its continuum limit and $r_s$ ceases 
to be a scaling parameter. This difference is shown in 
Fig. \ref{Fig1.a}, where one can see the universal scaling regime 
below $r_l^*(L)$ and the non universal lattice behaviors above 
$r_l^*(L)$. 

To show scaling when the GSs of different total momenta ${\bf K}$ 
correspond to a rigid molecule with ${\bf K}$-independent 
harmonic oscillations and a ${\bf K}$-dependent translation 
of the center of mass, we use the dimensionless quantum 
correction $F_0({\bf K},r_l)$ to the classical electrostatic 
energy: 
\begin{equation}
F_0({\bf K},r_l)=\frac{E_0({\bf K},r_l)-E_{cm}({\bf K})-E_{Coul}}
{E_0({\bf K}=0,U=0)} 
\label{F_0}.
\end{equation}
$F_0({\bf K},r_l)$ is shown in Fig. \ref{Fig1.a} as a function 
of $r_l=UL/t$ for different momenta ${\bf K}$ and $N=3$. One can 
see that the data obtained for $L=9$ and $18$ are on the same 
universal curve below $r_l^*(L)$, the curve depending on ${\bf K}$ 
below a value $r_l^W$ corresponding to $r_s^W \approx 50$. The 
universal  ${\bf K}$-independent scaling curve $0.2326 \sqrt{r_l}$ 
is obtained assuming that $E_0({\bf K}=0,U=0) \approx 8\pi^2 t/L^2$ 
as in the continuum limit:
\begin{eqnarray}
F_0 &=& {\sqrt{5}+\sqrt{2}\over 8 \pi^2} 
\sqrt{\sqrt{6}\pi^3\over 3} \sqrt{UL\over t} \nonumber \\
&=& {\sqrt{5}+\sqrt{2}\over \sqrt{96}} 
\left({18\over \pi}\right)^{1/4} \sqrt{r_s} \nonumber \\
&=& 0.5764 \sqrt{r_s} = 0.2326 \sqrt{r_l}.
\label{harm-theo-F0}
\end{eqnarray}
$F_0$ saturates above $r_l^*(L)$ to a value $4Nt/(8\pi^2 t/L^2)$ 
which is independent of $U$.

\section{Effects of a weak disorder in the lattice regime of 
an electron solid}
\label{section6}

\begin{figure}[ht]  
\vskip.2in
\centerline{\epsfxsize=8cm\epsffile{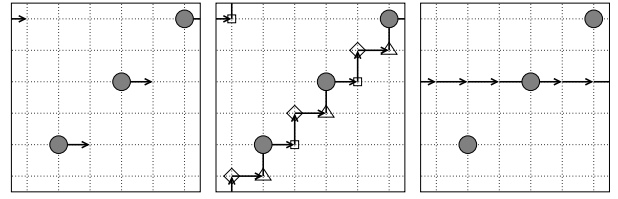}} 
\caption{Sketches of the hopping processes characterizing the 
three lattice regimes of a very weakly disordered lattice when 
$U>U^*$ for $L=6$ and $N=3$. From left to right: Ballistic Wigner 
molecule (BWM) when $U^*<U<U_{stripe}$; Coulomb Guided Stripe of 
Current (CGSC) when $U_{stripe}<U<U_{loc}$ and one particle motion 
out of a Localized Wigner Molecule (LWM) above $U_{loc}$.} 
\label{Fig5}
\end{figure} 

 Without interaction ($U=0$) one has three regimes (ballistic, 
diffusive and localized) as $W$ increases. Assuming Drude 
formula and Born approximation, the dimensionless conductance 
$g$ reads \cite{walker}: 
\begin{equation}
g \approx \frac{k_Fl}{2} \approx 96 \pi \frac{N}{L^2} \left(\frac{t}{W}
\right)^2. 
\end{equation}
For a Fermi momentum $k_F \approx \sqrt{4 \pi N}/(Ls)^2$, the elastic 
mean free path $l$ becomes
\begin{equation}
l \approx \frac{96 \sqrt{\pi} N}{L} \left (\frac{t}{W} \right)^2 s. 
\label{mean-free-path}
\end{equation}
If $U=0$, $N=3$, $L \approx 6 - 9$, this gives a ballistic motion 
when $W/t \approx 0.01 - 0.1$, a diffusive motion when $W/t \approx 1$, 
and strong Anderson localization for $W/t \approx 10 - 20$. 

 Let us first study the effect of a weak disorder in the simple 
lattice regime where the three particles form an electron solid 
(Wigner molecule). Since $r_l$ becomes an irrelevant parameter above 
$r_l^*$, we simply use the parameter $U$, $U^*$ being the interaction 
strength corresponding to $r_l^*(L)$ for the used values of $L$ and $t$. 
For $U> U^*$, the low energy physics of the $N$-body problem can be 
described by an effective one particle problem. For $N=3$ and $L/3$ 
integer, the Wigner molecule is oriented along one of the two diagonals 
of the square lattice. The average interparticle spacing 
$\left<d\right> = (\pm L/3,\pm L/3)$ and the fluctuations of $d$ are 
smaller than $s$ above $U^*$. Taking $t=1$, a fixed value of $L$ and a 
weak value of $W$, one gets three lattice regimes when one increases $U$ 
above $U^*$, characterized by three different lattice perturbation 
expansions. First, there is a Ballistic Wigner Molecule (BWM) on a scale 
$L$ smaller than the elastic mean free path $l_{cm}$ characterizing the 
motion of the center of mass. On this small scale, the disorder can be 
neglected, the motion of the center of mass remains ballistic and can be 
described by a non-random single particle lattice model with an effective 
nearest neighbor hopping term $t_{eff}\propto t^N/U^{N-1}$. Above an 
interaction $U_{stripe}>U^*$, the effects of disorder become relevant and 
must be included in the effective single particle model. This gives a new 
regime where the interplay between disorder and electronic correlations leads 
to a Coulomb Guided Stripe of Current (CGSC), since the current $I$ flows 
along the axis of the Wigner molecule. Eventually, the effective model 
breaks down when $U$ exceeds a large threshold $U_{loc}>U_{stripe}>U^*$, 
and we have the perturbative regime analyzed in refs. \cite{efros,sw} for 
a localized Wigner molecule (LWM). While the motion induced by $\Phi$ is 
along the shortest paths enclosing $\Phi$ in the BWM and in the LWM 
regimes, the motion in the intermediate CGSC regime uses a longer path of 
smaller electrostatic cost.   

In Fig. \ref{Fig5}, we have sketched the different hopping 
processes characterizing these three regimes when $L=6$. The left 
figure corresponds to a Ballistic Wigner molecule (BWM). The middle 
figure describes the  Coulomb Guided Stripe of Current (CGSC) induced 
by the interplay between disorder and electronic correlations. The right 
figure shows a Localized Wigner Molecule (LWM) carrying exponentially 
small one particle currents.

\subsection{Lattice regime for a Ballistic Wigner Molecule}
\label{subsection6.1}

 When $L$ is smaller than the elastic mean path $l_{cm}$ of the 
center of mass of the Wigner molecule, ${\bf K}$ does not vary 
and one can consider a subspace of total momentum ${\bf K}$. 
Assuming $L/3$ integer and $t/U \rightarrow 0$, the GS reads 
\begin{equation}
\left |\Psi_0({\bf K}) \right>= 
A \sum_{\bf j} e^{i{\bf K \cdot j}} C^{\dagger}_{\bf j} \left|0\right>,
\end{equation}
where 
\begin{equation}
C^{\dagger}_{\bf j}=c^{\dagger}_{\bf j} c^{\dagger}_{{\bf j}+(L/3,L/3)}
c^{\dagger}_{{\bf j}+(2L/3,2L/3)}
\end{equation}
and $A$ is a normalization constant. When $t=0$, we have $2L^2/3$ 
configurations $C^{\dagger}_{\bf j} \left|0\right>$ of identical 
Coulomb energy $E_{Coul}$. Without disorder, this large degeneracy 
is removed by a hopping term $t_{eff}\propto t^3/U^2$ coupling two 
nearest neighbor configurations $C^{\dagger}_{\bf j} \left|0\right>$  
and  $C^{\dagger}_{\bf j'} \left|0\right>$, as indicated in 
Fig. \ref{Fig5} left. At this order, one gets an effective one 
particle $L/3 \times L$ lattice model described by  
\begin{equation}
H_{eff}(W=0)=4Nt+E_{Coul}-t_{eff} \sum_{\left<{\bf j,j'}\right>} 
C^{\dagger}_{\bf j}C_{\bf j'}.
\label{effective-H}
\end{equation}
where the effective hopping term reads  
\begin{equation}
t_{eff}=\sum_{\{P\}} \frac{t^N}{\prod_{\gamma} 
\left(E_{Coul}-E_{P_\gamma} \right)}, 
\end{equation}
$P$ labeling a series of intermediate configurations coupling 
${\bf j}$ and ${\bf j'}$ of Coulomb energy $E_{P_\gamma}$.
This gives 
\begin{equation}
t_{eff} \sim {t^N \over U^{N-1}}L^{3N-3}, 
\end{equation}
and for $N=3$
\begin{equation}
t_{eff}= 6 {t^3 \over \left({7\sqrt{6} \over 36}
{U \pi\over L^3}\right)^2 }.
\label{hop-lat-clean}
\end{equation}

 Including the flux in the longitudinal hopping terms  
$t_{eff} \exp \left(i 2\pi N\Phi / \Phi_0 L\right)$, 
one gets for the change of the GS energy 
\begin{equation}
\Delta E_0\left(\Phi,{\bf K}\right) = 2t_{eff} \left(\cos(K_x+
\frac{2\pi N\Phi}{\Phi_0 L})-\cos K_x \right)
\label{delta-BWM}.
\end{equation}
 
\begin{figure}[ht]  
\vskip.2in
\centerline{\epsfxsize=8cm\epsffile{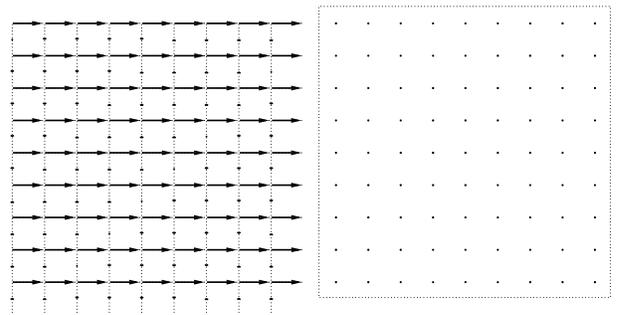}} 
\caption{Ballistic translation of a Wigner Molecule in the 
lattice regime (BWM regime), obtained in a disordered sample 
with $N=3$, $L=9$, $W=0.01$, $t=1$ and $U=300$. Left figure: 
map of currents for $\Phi=0.05 \Phi_0$. Right figure: corresponding 
site occupation numbers $n_{\bf j}^0$.} 
\label{Fig6a}
\end{figure}

In Fig. \ref{Fig6a} as in the following ones, the maps 
of local currents characterizing a disordered sample 
and the corresponding site occupation numbers $n_{\bf j}^0$ 
given by 
\begin{equation}
n_{\bf j}^0= \left<\Psi_0(U)\left|c^{\dagger}_{\bf j}c_{\bf j} 
\right| \Psi_0(U)\right>
\end{equation}
are shown using the same convention: the arrows between 
neighboring sites give the local currents, the largest local 
current found in the sample is shown by an arrow of length 
$s$, while $n_{\bf j}^0=1$ is shown by a disk of diameter $s$. 
A small value $\Phi=0.05 \Phi_0$ is used for driving the 
persistent currents.

 In Fig. \ref{Fig6a}, the map of local currents are given for 
a weakly disordered sample ($W=0.01$ and $t=1$) with an interaction 
strength $U=300$ above $U^*$ and the  corresponding site occupation numbers 
$n_{\bf j}^0$. The map of currents and the density exhibit an 
almost perfect translational invariance in this ballistic regime 
and gives an illustration of the BWM motion characterized by 
$t_{eff}$. 

\subsection{Lattice regime for a Coulomb Guided Stripe of Current}
\label{subsection6.2}
  
\begin{figure}[ht]  
\vskip.2in
\centerline{\epsfxsize=8cm\epsffile{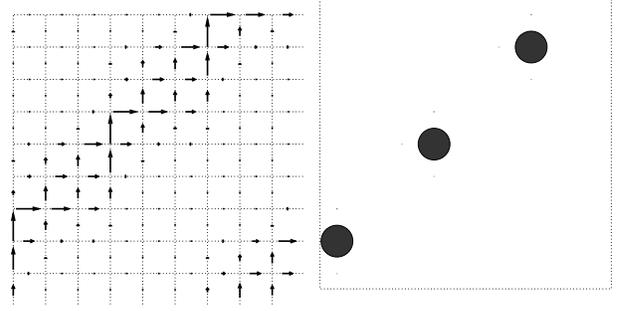}} 
\caption{Coulomb Guided Stripe of Current (CGSC) 
in the lattice regime, obtained in a disordered sample with 
$N=3$, $L=9$, $W=t=1$ and $U=1000$. Left figure: map of 
currents for $\Phi=0.05 \Phi_0$. Right figure: corresponding site 
occupation numbers $n_{\bf j}^0$.} 
\label{Fig6b}
\end{figure}

  As one continues to increase $U$, the effect of disorder cannot be 
ignored and the Wigner molecule finishes to be pinned in the GS 
configuration $C^{\dagger}_{\bf j} \left|0\right>$ of minimum 
Coulomb energy for which $\sum_{\bf j} \epsilon_{\bf j}$ is also minimum. 
As one can see in  Fig. \ref{Fig6b} (right), a complete localization 
of the center of mass of the rigid molecule is achieved for a disordered 
sample with $W=t=1$ and $U=1000$. The GS probability to occupy a site 
is almost zero outside three sites where $C^{\dagger}_{\bf j} \left|0\right>$ 
has a minimum energy. It is likely that this complete localization is 
obtained after a small diffusive regime. In this pinned regime, two 
competing motions are now relevant for enclosing the flux, which are 
sketched in Fig. \ref{Fig5} (middle and right). The CGSC motion 
(Fig. \ref{Fig5} middle) is larger when $U_{stripe}<U<U_{loc}$ while 
the LWM motion (Fig. \ref{Fig5} right) dominates the longitudinal motion 
when $U_{loc}<U$. Let us first study the CGSC motion. The effect of 
disorder can be included by adding random site potentials 
$W \omega_{\bf j}$ to the effective one particle model (\ref{effective-H}):
\begin{equation}
H_{eff}(W)= H_{eff}(W=0)+W \sum_{\bf j} \omega_{\bf j}C^{\dagger}_{\bf j}
C_{\bf j},
\label{Effective-H-dis}
\end{equation}
where $\omega_{\bf j}=\epsilon_{\bf j}+\epsilon_{{\bf j}+(L/3,L/3)} + 
\epsilon_{{\bf j}-(L/3,L/3)}$. 

Assuming the Hamiltonian (\ref{Effective-H-dis}), the GS flux dependence 
is obtained at the order $2L/3$ of an expansion in powers of $t_{eff}/W$, 
corresponding to the $(L/3,L/3)$-translation shown in Fig. \ref{Fig5} 
middle. One gets for the GS energy change
\begin{eqnarray}
\Delta E_0\left(\Phi\right) &\propto&
\frac{t_{eff}^{2L/3} \cos\left( 2 \pi \Phi/\Phi_0  \right)}
{{{\sqrt N} W}^{2L/3-1}} \\ 
&\propto& \frac{t^{2L} L^{4L} \cos\left( 2 \pi \Phi/\Phi_0\right)}
{W^{2L/3-1} U^{4L/3}}, 
\label{delta-CGSC}
\end{eqnarray}
which gives for the two sizes $L$ which we have studied
\begin{eqnarray}
\Delta E_0\left(L=6,\Phi\right) &\propto& \frac
{t^{12} } {W^{3} U^{8}} \cos\left( 2 \pi \Phi/\Phi_0\right) \\  
\Delta E_0\left(L=9,\Phi\right) &\propto& \frac
{t^{18}} {W^{5} U^{12}} \cos\left( 2 \pi \Phi/\Phi_0\right) 
\end{eqnarray}
respectively. 

The current $I$ forms a stripe which follows the axis of the 
pinned Wigner molecule for enclosing the flux. The path is longer 
than the shortest path, but it uses intermediate configurations of 
minimum Coulomb energy. An illustration of such a stripe is given 
in Fig. \ref{Fig6b} (left), obtained using a disordered sample 
where $W=t=1$ and $U=1000$: the pinned Wigner molecule carries in the 
CGSC regime a current of components $I_l=I_t$, in contrast to the BWM 
and LWM regimes where $I_t$ is negligible compared to $I_l$. 

\subsection{Lattice regime for a Localized Wigner Molecule}
\label{subsection6.3}

\begin{figure}[ht]  
\vskip.2in
\centerline{\epsfxsize=8cm\epsffile{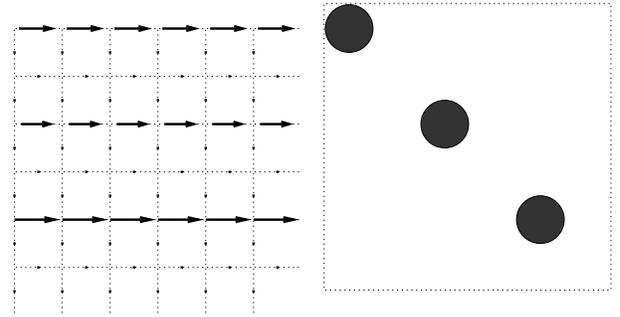}} 
\caption{One particle motions out of a Localized Wigner Molecule (LWM) 
in the lattice regime: Disordered sample with $N=3$, $L=9$, $t=1$, 
$W=20$ and $U=1000$. Left figure: map of currents for $\Phi=0.05 
\Phi_0$. Right figure: corresponding site occupation numbers 
$n_{\bf j}^0$.} 
\label{Fig6c}
\end{figure}

 When $U$ becomes very large, the previous CGSC remains and gives 
the transverse current $I_{t}$, but the main contribution to $I_{l}$
is no longer given by the effective Hamiltonian (\ref{Effective-H-dis}). 
This is because one particle hops, using $L-1$ intermediate states of 
high Coulomb energy become more advantageous than the correlated 
effective hopping using $2L-1$ states of smaller Coulomb energy. 
For $N=3$, this corresponds to one particle hops along the shortest path 
inclosing the flux (see Fig. \ref{Fig5} right). To numerically get this 
regime for a very weak disorder and $L=9$ requires a very large value 
of $U/t$, where the numerical results become inaccurate. Nevertheless, 
an illustration of this regime using a sample with a large disorder 
($W/t=20$) is given for $U/t=1000$ in Fig. \ref{Fig6c}, with the 
corresponding GS occupation numbers.  

The GS energy change $\Delta E_0\left(\Phi\right)$ reads:
\begin{equation}
\Delta E_0\left(\Phi\right) \propto \frac
{t^L L^{3L-3}}
{U^{L-1}} \cos\left( 2 \pi \Phi/\Phi_0  \right)
\label{delta-LWM}
\end{equation}
which gives for the two sizes $L$ which we have studied
\begin{eqnarray}
\Delta E_0 \left(L=6,\Phi\right) &\propto& \frac
{t^6}{U^5} \cos \left( 2 \pi \Phi/\Phi_0  \right)   \\
\Delta E_0 \left(L=9,\Phi\right) &\propto& \frac
{t^9}{U^8} \cos \left( 2 \pi \Phi/\Phi_0\right) 
\end{eqnarray}
respectively. 

\subsection{Crossovers between the different lattice regimes}
\label{subsection6.4}

 To obtain the threshold values where a crossover between 
different lattice regimes occurs, we compare the GS energy changes 
$\Delta E_0\left(\Phi\right)$ of the different lattice regimes 
(Eqs. \ref{delta-BWM}, \ref{delta-CGSC} and \ref{delta-LWM}).

Increasing $W/t$ for a ratio $U/t>U^*/t$, the BWM-CGSC crossover 
takes place when $W$ reaches a threshold value 
$W_{stripe} \approx t_{eff}$. This gives: 
\begin{equation}
\frac{W_{stripe}}{t} \propto \left(\frac{t}{U}\right)^2 L^6.
\end{equation}
The CGSC-LWM crossover takes place when $W/t$ reaches a second 
threshold 
\begin{equation}
\frac{W_{loc}}{t} \propto \left(\frac{t}{U}\right)^{\frac{L+3}{2L-3}} 
L^{\frac{3L+9}{2L-3}}
\end{equation}

If one increases $U/t$ above $U^*/t$ for a given ratio $W/t$,  
the BWM-CGSC crossover occurs when $U/t$ becomes of the order of       
\begin{equation}
\frac{U_{stripe}}{t} \propto \left(\frac{t}{W}\right)^{1/2} L^3,
\label{threshold-stripe}
\end{equation}
followed by the CGSC-LWM crossover when $U/t$ reaches 
\begin{equation}
\frac{U_{loc}}{t} \propto \left(\frac{t}{W}\right)^{\frac{2L-3}{L+3}} 
L^{\frac{3L+9}{L+3}}.
\label{threshold-loc} 
\end{equation}

\begin{figure}[ht]  
\vskip.2in
\centerline{\epsfxsize=8cm\epsffile{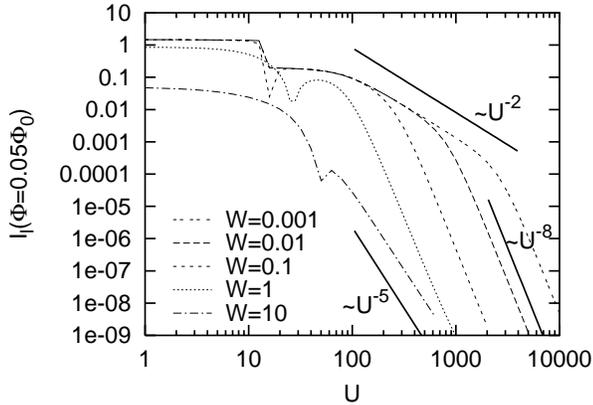}} 
\caption{$L=6$ and $t=1$: Amplitude $|I_l(\Phi=0.05 \Phi_0)|$ 
of the total longitudinal current as a function of $U$ 
for a disordered sample with increasing values of $W$. The thick 
solid lines give the expected lattice decays.} 
\label{Fig7}
\end{figure} 

\begin{figure}[ht]  
\vskip.2in
\centerline{\epsfxsize=8cm\epsffile{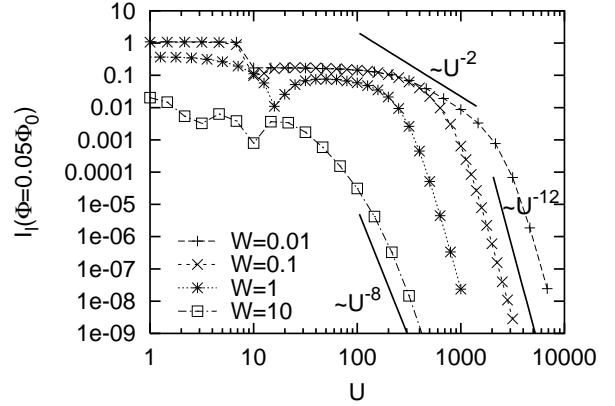}} 
\caption{$L=9$ and $t=1$: Amplitude $|I_l(\Phi=0.05 \Phi_0)|$ 
of the total longitudinal current as a function of $U$ for a 
disordered sample with increasing values of $W$. The thick solid 
lines give the expected lattice decays.} 
\label{Fig7a}
\end{figure} 

\section{Numerical results for $N=3$ particles.}
\label{section7}

The study of different samples confirms the dependences 
of the effective hopping terms predicted for the three 
lattice regimes as a function of the lattice parameters $U$, $W$ 
and $t$ in the expected ranges of parameters. We show numerical 
results for the total currents $I_l$ and $I_t$, and their ratios 
$I_t/I_l$ obtained using disordered samples, fixing two parameters 
($U$, $W$ or $t$) and varying the third, when a small flux 
$\Phi=0.05 \Phi_0$ is enclosed. We give the typical behavior obtained 
using a single sample, without ensemble average. The sample to sample 
fluctuations are negligible for weak disorders, and can be more important 
when $W$ becomes large.  

\subsection{Persistent currents as a function of $U$}
\label{subsection7.1}

 In Figs. \ref{Fig7} and \ref{Fig7a}, the amplitude of the 
total longitudinal current $|I_l|$ is shown as a function of 
$U$ for $t=1$ and various values of $W$, for $L=6$ and $9$ 
respectively. The power laws with the predicted exponents 
characterize the decay of the currents ($U^{-2},U^{-4L/3}$ 
and $U^{1-L}$) in the three expected lattice regimes. 
\begin{figure}[ht]  
\vskip.2in
\centerline{\epsfxsize=8cm\epsffile{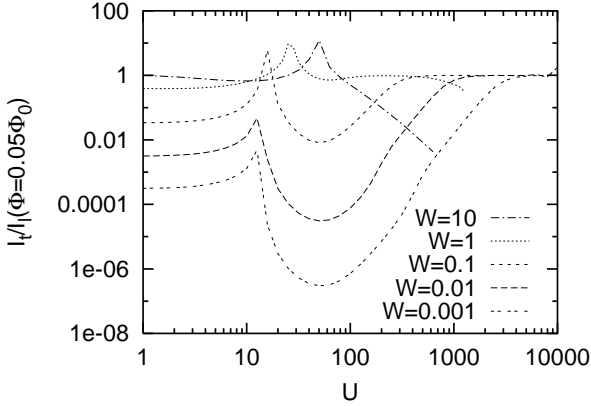}} 
\caption{$L=6$ and $t=1$: Ratio $I_t/I_l$ for $\Phi=0.05 \Phi_0)$ as a 
function of $U$ for increasing values of $W$.} 
\label{Fig8}
\end{figure} 

\begin{figure}[ht]  
\vskip.2in
\centerline{\epsfxsize=8cm\epsffile{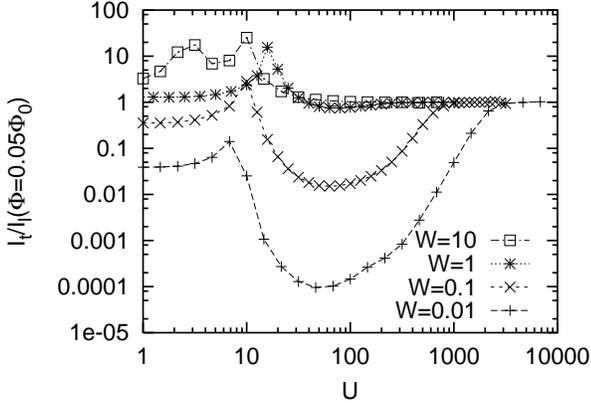}} 
\caption{$L=9$ and $t=1$: Ratio $I_t/I_l$ for $\Phi=0.05 \Phi_0$ as a 
function of $U$ and increasing values of $W$.} 
\label{Fig8a}
\end{figure} 

 In Figs. \ref{Fig8} and \ref{Fig8a}, the ratio $|I_t/I_l|$ 
is given as a function of $U$. Since the stripe of current 
characterizing the CGSC-regime yields a ratio $|I_t/I_l|=1$, 
one can see on those figures the values of $U$ where the 
CGSC-regime takes place. For small values of $U$ which are 
below the lattice threshold $U^*$, one can see in Figs. \ref{Fig7} 
and \ref{Fig7a} that $|I_l|$ and $|I_t|$ are essentially independent 
of $U$, excepted around a value $U_{lc}$ where an avoided level 
crossing occurs. The value of $U_{lc}$ depends on $W$ and $t$ 
and exhibits small sample to sample fluctuations. This avoided 
crossing yields a change of the sign of $I_l$, a sharp drop 
of $|I_l|$ and a singularity in $|I_t/I_l|$ which can be seen 
in  Figs. \ref{Fig8} and \ref{Fig8a}. The continuum behaviors 
taking place when $U<U^*$, where $I_l$ and $I_t$ are essentially 
independent of $U$ instead of exhibiting the lattice decays while 
$|I_t/I_l| \approx 1$, are studied in section \ref{section9}. 

\subsection{Persistent currents as a function of $W$}
\label{subsection7.2}

\begin{figure}[ht]  
\vskip.2in
\centerline{\epsfxsize=8cm\epsffile{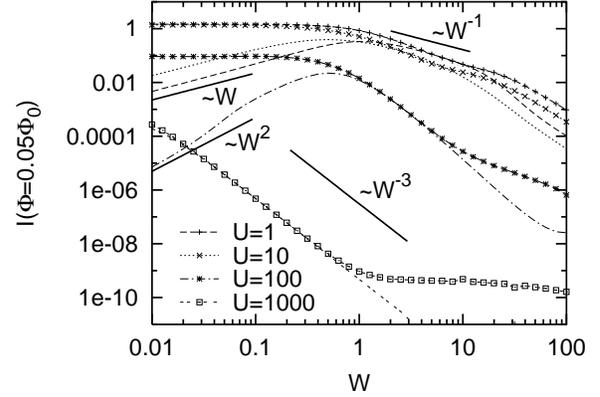}} 
\caption{$L=6$ and $t=1$: Amplitudes $|I_l|$ (indicated by a line 
with a symbol) and $|I_t|$ (indicated by a line without symbol) 
of the total longitudinal and transverse currents as a function 
of $W$. The data are obtained using a disordered sample with 
$\Phi=0.05 \Phi_0$. The symbols and lines corresponding to different 
values of $U$ are defined in the figure. The thick solid lines give 
the $W^{-3}, W^{-1}, W^{1}$ and $W^{2}$ decays respectively.} 
\label{Fig9}
\end{figure} 

\begin{figure}[ht]  
\vskip.2in
\centerline{\epsfxsize=8cm\epsffile{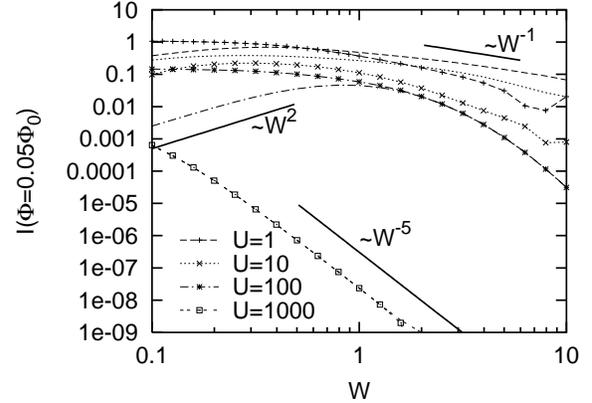}} 
\caption{$L=9$ and $t=1$:  Amplitudes $I_l$ (indicated by a line 
with a symbol) and $I_t$ (indicated by a line without symbol) 
as a function of $W$. The studied values of $U$ are represented as 
in Fig.\ref{Fig9}. The thick solid lines give the $W^{-5}, W^{-1}$ 
and $W^{2}$ decays respectively.} 
\label{Fig9a}
\end{figure} 

 Figs. \ref{Fig9} and \ref{Fig9a} show for $L=6$ and $9$ 
respectively how the amplitudes $|I_l|$ (indicated by lines 
with symbols) and $|I_t|$ (indicated by the same lines 
without symbol) vary as a function of $W$ for $t=1$ and various 
values of $U$. When $U$ is large ($U=100 - 1000$), one can see 
the expected lattice behaviors: the  CGSC-decay $I_l \propto 
W^{-(2L/3-1)}$ for $W_{stripe}<W<W_{loc}$ followed above 
$W_{loc}$ by the LWM-regime where $I_l$ saturates to a small 
value independent of $W$. When $U$ is smaller ($U=1 - 10$), 
$I_l$ is first almost independent of $W$, before decaying as 
$W^{-1}$ when $1 < W <10$ for $L=6$ and $9$. 

The transverse current $I_t$ begins to increase as a function 
of $W$ from the zero value characterizing the limit $W=0$. 
This increase of $I_t$ takes place for the values of $W$ where 
$I_l$ is roughly independent of $W$. This increase goes as $W^2$ 
for large interactions  ($U=100 - 1000$) and as $W$ for small 
interactions ($U=1 - 10$), as underlined in Fig.\ref{Fig9}. These 
two different powers can be explained by a perturbative expansion 
in powers of $W$. When $W=0$, GSs of different total momenta 
${\bf K}$ are degenerate when $U$ is small and $I_t \propto W$. 
As $U$ increases, there is a level crossing for $W=0$ above 
which there is a non-degenerate GS of momentum ${\bf K}=0$ and 
$I_t \propto W^2$  when $U$ is large. The increase of $I_t$ is 
followed by a regime where $I_t \approx I_l$ when $L=6$ or exhibits 
the same decays ($W^{-1}$ or $W^{-(2L/3-1})$) than $I_l$ when $L=9$. 
For large $U$ and above $W_{loc}$, $I_t$ continues to decay as in 
the lattice CGSC-regime ($I_t \propto W^{-(2L/3-1)}$) while $I_l$ 
saturates to a $W$-independent value. 

\subsection{Persistent currents as a function of $t$}
\label{subsection7.3}

\begin{figure}[ht]  
\vskip.2in
\centerline{\epsfxsize=8cm\epsffile{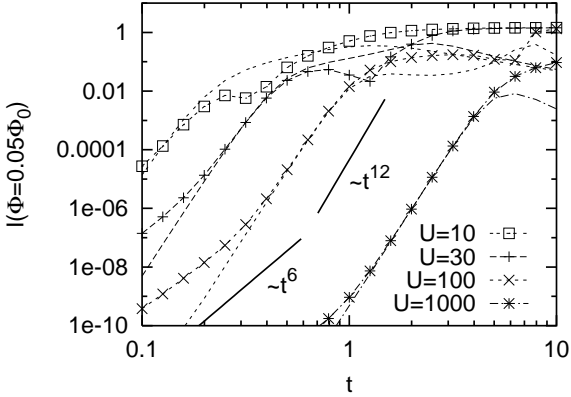}} 
\caption{$L=6$ and $W=1$: Amplitudes $|I_l|$ (lines with symbols) 
and $|I_t|$ (lines without symbol) as a function of $t$, obtained 
using a disordered sample with increasing values of $U$ and 
$\Phi=0.05 \Phi_0$. The thick solid lines give the expected lattice 
behaviors.} 
\label{Fig11}
\end{figure} 

\begin{figure}[ht]  
\vskip.2in
\centerline{\epsfxsize=8cm\epsffile{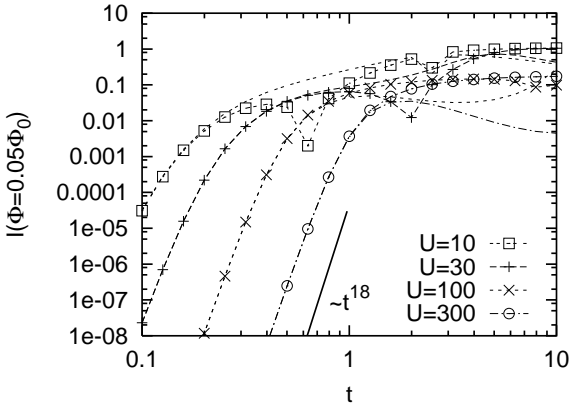}} 
\caption{$L=9$ and $W=1$: Amplitudes $|I_l|$ (line with symbols) 
and $|I_t|$ (line without symbol) as a function of $t$, obtained 
using a disordered sample with increasing values of $U$ and 
$\Phi=0.05 \Phi_0$. The thick solid lines give the expected 
lattice behaviors.} 
\label{Fig11a}
\end{figure} 

In Figs. \ref{Fig11} and \ref{Fig11a}, the amplitudes $|I_l|$ 
and $|I_t|$ are shown as a function of $t$ for $W=1$ and various 
values of $U$, using disordered samples of size $L=6$ and $9$ 
respectively. The power laws shown by thick lines give the 
decay of the currents $t^{L},t^{2L}$ for $L=6$ and $9$ 
respectively, as expected for lattice regimes ($t<t^*$) with 
$W=1$. The $t^3$-behavior expected in the BWM-regime requires a 
smaller disorder to be observed. For $t>t^*$, the behaviors are 
similar to those observed for $t=1$ and $U<U^*$. In the lattice 
LWM regime ($t \rightarrow 0$), one can see that  $|I_t|$ 
continues to exhibit the lattice CGSC-behavior $\propto t^{2L}$ 
while $|I_l| \propto t^{L}$ becomes much larger. 

\section{Continuum-Lattice crossovers with disorder} 
\label{section8}

We have shown in section \ref{section4} that $I \propto 
\left<{\bf K}\right>$ unless the GS begins to occupy states of high 
momenta ${\bf k}$. Therefore, the breakdown of the interaction-invariance 
of $I$ is a consequence of a localization-delocalization transition 
induced by $U$ or $W$ in momentum space, unless one has a level crossing. 
Without disorder, a jump in the GS total momentum ${\bf K}$ 
can be induced by level crossings between GSs of different ${\bf K}$. 
In the presence of disorder, these level crossings become avoided 
level crossings. We discuss in the following sub-section how avoided 
level crossings limit the interaction-invariance of $I$, outside the 
lattice regimes where the relation $I \propto \left<{\bf K}\right>$ is 
broken.

\subsection{Level crossings below $r_s^W$ and momentum conservation}
\label{section8.1}

\begin{figure}[ht]  
\vskip.2in
\centerline{\epsfxsize=8cm\epsffile{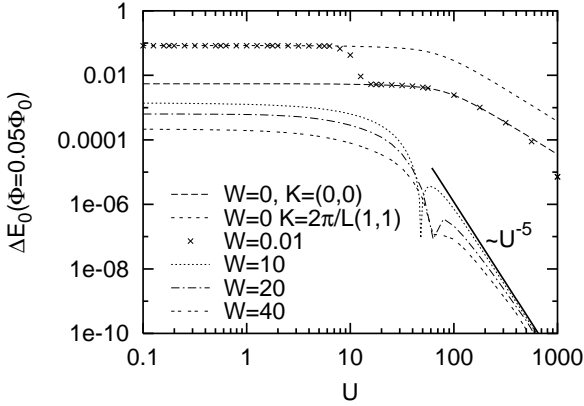}} 
\caption{$L=6$ and $t=1$: Change $\Delta E_0$ of 
the GS energy induced by $\Phi=0.005 \Phi_0$ as a function 
of $U$ for different values of $W$.} 
\label{Fig4}
\end{figure}

  When $W=0$, ${\bf K}$ is conserved and one can follow the 
GS of given ${\bf K}$ as a function of $U$. When $U=0$, the 
GS momentum is ${\bf K}=0$ only if the Fermi shell is full. 
For a square lattice, this corresponds to $N=1,5,9,\ldots$. 
For $U^W <U < U^*$, the GS is a Wigner molecule with harmonic 
oscillations and a delocalized center of mass. This is what we 
have shown for $N=3$ in section. \ref{section5}. The GS energy 
\begin{equation}
E_0({\bf K},r_s) \approx E_{cm}({\bf K})+ E_{rel}(r_s)+E_{coul}
\end{equation}
depends on ${\bf K}$ via  the motion of the center of mass only.
In this continuum Wigner regime, the kinetic energy of the center 
of mass is minimum for ${\bf K}=0$. 
So for $N=1,5,9,\ldots$, the GS remains in the sub-space of 
${\bf K}=0$ when $U$ varies and does not exhibit level crossings 
between GSs of different ${\bf K}$. For $N=3$, this is different, 
because the  GS momentum ${\bf K} \neq 0$ at $U=0$ and must be zero 
above $U^W$. This yields a GS-level crossing at an interaction 
$U_{lc}$ which is necessary smaller than $U^W$. In the presence of 
a weak disorder, this crossing becomes weakly avoided. Following the 
true GS as a function of $U$, we expect that its total momentum 
will be conserved, outside the avoided crossing where ${\bf K}$ 
exhibits a sharp drop from a value ${\bf K}\neq 0$ to ${\bf K}=0$. 
This is illustrated in Fig. \ref{Fig4} where one can see the change 
$\Delta E_0$ of the GS energy induced by $\Phi=0.05 \Phi_0$ 
as a function of $U$. $\Delta E_0\left({\bf K}\right)$ are given when 
$W=0$ and $L=6$ for ${\bf K}=(0,0)$ and $2 \pi /L (1,1)$. One can see 
the interaction independent behaviors up to a lattice threshold 
$U^*(L=6) \approx 100$, followed by the lattice decay $\propto U^{-2}$. 
When we include a very weak disorder ($W/t=0.01$), and follow the true  
GS,  $\Delta E_0 (W=0.01) \approx \Delta E_0\left(W=0,{\bf K}=2 
\pi /L (1,1)\right)$ up to the avoided crossing taking place at an 
interaction $U_{lc}\approx 10$ where $\Delta E_0$ jumps towards the 
GS-behavior characterizing $W=0$ and ${\bf K}=2 \pi /L (0,0)$.  
 
 We now study when the relation $I \propto \left<{\bf K}\right>$ 
is broken in a disordered lattice of elastic mean free path $l$ and 
localization length $\xi$ without interaction. To obtain the threshold values 
where the lattice-continuum crossover takes place, we use the change 
$\Delta E_0$ of the GS energy instead of $I$, and look when the 
$\Delta E_0(U)$ characterizing the three lattice regimes described in 
section \ref{section6} become of the order of $\Delta E_0(U=0)$.

\subsection{Lattice-Continuum crossover in the ballistic regime 
($L< l$)}
\label{subsection8.2}

 When $L<l$, $\Delta E_0 \approx t$ without interaction and goes as 
$t_{eff}$ in the BWM-regime. The condition:  
\begin{equation}
t \approx \frac{L^6t^3}{U^2},
\label{threshold-bal}
\end{equation}
gives for the threshold $U^*/t$ the same equation   
\begin{equation}
\frac{U^*}{t} \approx L^3,  
\end{equation}
than Eq. \ref{criterion4-bis}.

\subsection{Continuum-Lattice crossover in the diffusive regime 
($l<L<\xi$)}
\label{subsection8.3}

 When the one particle motion is diffusive without interaction, 
$\Delta E_0(U=0) \propto I(U=0) \propto \hbar e k_Fl /(m Ls^2) 
\propto g t /L^2$ (see refs \cite{Montambaux-Bouchiat,Cheung-Gefen}). 
When $\Delta E_0$ becomes smaller than $\Delta E_0(U=0)$ for values 
of $U>U_{stripe}$, the exponent of the decay which can be seen in 
Figs. \ref{Fig7} and \ref{Fig7a} (curves for $W=0.1t$ and $1t$) shows 
that one enters in the lattice CGSC-regime. Since $g \propto (t/(LW))^2$, 
the crossover threshold $U_{stripe}$ is given by the condition 
$\Delta E_0(U=0)\approx gt/L^2 \approx \Delta E_{0,stripe}$, or 
\begin{equation}
\frac{t^3}{W^2L^4} \approx \frac{t^{2L} L^{4L}}{W^{2L/3-1} U^{4L/3}},
\end{equation}
which gives a continuum-lattice crossover taking place in the 
diffusive regime when:
\begin{equation}
\frac{U_{stripe}}{t} \propto \left(\frac{t}{W}\right)^{1/2} L^3.
\label{threshold-dif-2}
\end{equation}
Comparing Eq. \ref{threshold-dif-2} and Eq. \ref{threshold-stripe}, 
one can see that one gets the same relation for the continuum-lattice 
crossover in a diffusive sample (e.g. $W \approx t$ for $L=6 - 9$) 
and for the BWM-CGSC crossover in the lattice regime of a 
weakly disordered sample (e.g. $W \approx 0.01 t$ for $L= 6 - 9$). 

\subsection{Continuum-Lattice crossover in the localized regime 
($L> \xi$)}
\label{subsection8.4}

 When all the particles are localized without interaction, 
$\Delta E_0 \approx t \exp -(L/\xi)$ where the localization 
length $\xi \approx \sqrt{t/W}$ for a large ratio $W/t$ 
(see ref.\cite{kroha}). When $\Delta E_0$ decays as a 
function of $U$, we can see in Fig. \ref{Fig4} that this 
decay corresponds to the lattice LWM-regime. The continuum-lattice 
crossover in strongly disordered samples takes place when 
\begin{equation}
t \exp \left(- L \sqrt{\frac{W}{t}}\right) \approx 
\frac{t^{L} L^{3L-3}}{U^{L-1}}
\end{equation}
which gives an interaction threshold  
\begin{equation}
\frac{U_{glass}}{t}\approx L^3 \exp \left(\sqrt{\frac{W}{t}}\right)
\end{equation}
above which one has the LWM-regime instead of a correlated Anderson 
insulator (CAI) for a large size $L$. 

Let us mention that a complete discussion of the various possible 
lattice regimes requires to consider also the lattice regimes yielded 
by large values of $W$ as we have considered the lattice regimes 
yielded by large values of $U$, since the GS can occupy large momenta 
${\bf k}$ either when $U$ or $W$ are large. A complete study will 
require also to determine if the CAI regime is in a continuum limit 
or in a lattice limit, as a function of the disorder strength $W$ and 
$L$. We do not discuss in more details this issue, though the data 
which we show for large values of $W/t$ are certainly not in a continuum 
regime. 

\subsection{Sketch of the phase diagram of the different lattice 
regimes obtained for $N=3$}
\label{subsection8.5}

\begin{figure}[ht]  
\vskip.2in
\centerline{\epsfxsize=8cm\epsffile{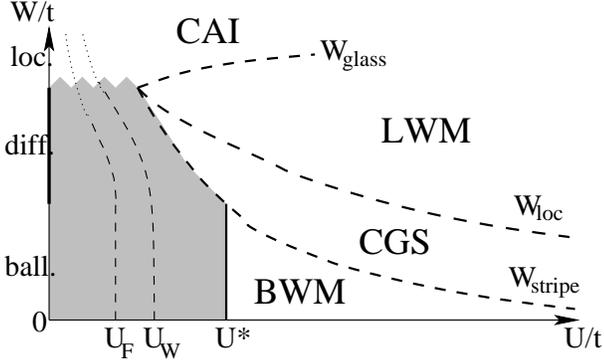}} 
\caption{Different lattice regimes for $N=3$ spinless 
fermions in $L \times L$ disordered lattices. The shaded 
part of the $\left(U/t,W/t\right)$ plane gives the continuum 
regimes where $I$ does not depend on $U/t$. The thick dashed 
lines give the threshold values $W_{stripe}$ and $W_{loc}$
and $W_{glass}$ separating different lattice regimes.
$U^*$, $U_W$ and $U_F$ are defined in section \ref{section3}.
The continuum and lattice CAI regimes occuring at large disorders 
are not separated.} 
\label{Fig10}
\end{figure} 

 We have sketched in Fig.\ref{Fig10} the different lattice regimes 
characterizing $N=3$ spinless fermions in the plane 
$\left(U/t,W/t\right)$. The shaded part of the $\left(U/t,W/t\right)$ 
plane gives the continuum regimes which we will study in more details 
in section \ref{section9}. The non shaded part gives the lattice 
regimes which we have studied. Let us summarize the behaviors 
obtained by increasing $U$ for small disorder strengths $W$. 

When $W=0$, one gets between $U_F$ and $U_W$ a continuum 
supersolid molecule between the Fermi system and the continuum 
Wigner molecule before having a lattice Wigner molecule above 
$U^*$ (see section \ref{section5}). 

When $W/t < 1$ and $L\approx 6 - 9$, the continuum ballistic 
regimes are followed by three lattice regimes taking place above 
$U^*/t$: a BWM regime for $U^*<U< U_{stripe}$; a CGSC regime 
for $U_{stripe}<U< U_{loc}$ and eventually a LWM regime above 
$U_{loc}$. The disorder is irrelevant below $U_{stripe}$, unless 
the system is at the vicinity of an avoided level crossing. 

When $W/t$ is large enough to yield Anderson localization inside a 
small scale $L \approx 6 - 9$, the Fermi glass with Anderson localization 
(CAI regime) becomes a highly correlated solid as $U$ increases, 
through an intermediate regime studied in Ref. \cite{bwp1}. Three 
examples are given in Fig. \ref{Fig4} for $W/t=10,20,40$, showing 
that when $U$ exceeds $U_{glass}$, a localized regime where $I$ is 
independent of $U$ gives rise to another localized regime where $I$ 
becomes independent of $W$. In those three cases, $I_l$ changes 
its sign around $U_{glass}$. As pointed out in Refs. \cite{bwp1,bwp2}, 
the sign of $I_l$ fluctuates from sample to sample for weak $U$ while 
it becomes non-random at large $U$.

We study in the following section the effect of $U$ upon the continuum 
limit of a diffusive sample where $W=t=1$ for $L=6$ and $9$.

\section{Analysis of the continuum regimes of a diffusive sample}
\label{section9}

 For a ratio $W/t \approx 1$, there is a diffusive motion when $U/t=0$ 
and $L \approx 6$ or $9$. We study the continuum limit where $I$ does 
not exhibit the power law decays characteristic of lattice regimes. 
For values of $U/t$ smaller than the lattice threshold $U^*/t$, one 
can indeed see in Figs. \ref{Fig7} and \ref{Fig7a} that $|I_l|$ and 
$|I_t|$ are almost independent of $U$, excepted around the value 
$U_{lc}$ where an avoided level crossing occurs. $U_{lc}$ depends on 
$W$ and $t$ and exhibits sample to sample fluctuations. This 
avoided crossing yields a change of the sign of $I_l$, a sharp drop 
of $|I_l|$ and singularities in $|I_t/I_l|$ (see Figs. \ref{Fig8} and 
\ref{Fig8a}). Since the total current $I$ is independent of $U/t$ 
outside level crossings in the continuum limit, one needs to study the 
local currents and the corresponding particle densities for detecting 
the different continuum regimes.

\subsection{Continuum regime for a Coulomb Guided Stripe of Current}
\label{subsection9.1}

\begin{figure}[ht]  
\vskip.2in
\centerline{\epsfxsize=8cm\epsffile{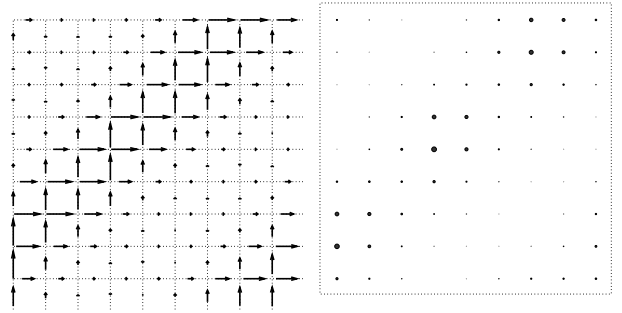}} 
\caption{Coulomb guided stripe of current in the continuum 
limit. Disordered sample with $N=3$, $L=9$, $W=1$, $t=1$ 
and $U=50$. Left figure: map of currents for $\Phi=0.05 
\Phi_0$. Right figure: corresponding site occupation numbers 
$n_{\bf j}^0$.} 
\label{Fig12}
\end{figure}

As one can see in Figs. \ref{Fig8} and \ref{Fig8a}, $|I_t/I_l|=1$ 
below $U^*/t$, while $I_l$ and $I_t$ are independent of $U$ instead 
of exhibiting the lattice decays. This shows us that the regime of 
current stripes is robust, and persists outside the lattice limit, in 
the continuum diffusive limit. This continuum CGSC regime is illustrated 
in Fig. \ref{Fig12}. The left figure shows the map of local currents 
obtained in a disordered sample with $N=3$, $L=9$, $W=1$, $t=1$ 
and $U=50$. The current exhibits the Coulomb guided flow along the diagonal 
direction, while one can see in Fig. \ref{Fig7a} that $I$ does not vary 
as a function of $U$ around $U=50$. The corresponding site occupation 
numbers given in Fig. \ref{Fig12} (right) form an extended diagonal stripe, 
instead of being localized upon three main sites, as in the CGSC lattice 
regime shown in Fig. \ref{Fig6b} (right). 

\subsection{Diamagnetic or Paramagnetic currents?}
\label{subsection9.3}

 Before considering weaker values of $U$ to be closer to the 
quantum Fermi limit, it is useful to mention a theorem proved 
by Leggett \cite{leggett1} for one dimensional spinless fermions with 
arbitrary interaction and external potentials. This theorem 
allows to determine the parity of the number of particles 
carrying $I$ from the sign of $I$. Using a certain variational 
ansatz for the GS wave-function, one can show that $I$ is 
diamagnetic for an odd number $N$ of particles when 
$\Phi \approx n\Phi_0$ or for an even number $N$ of particles 
when $\Phi \approx (n+1/2)\Phi_0$, where $n$ 
is an integer. This means that the charge stiffness 
\begin{equation}
D=\frac{(-1)^N L}{2} \left(E_0(0)-E_0(\pi)\right) \geq 0
\label{Leggett}
\end{equation}
in one dimension. To odd numbers $N$ correspond  diamagnetic currents, 
while they are paramagnetic if $N$ is even. This result has always been 
verified \cite{sjwp} in numerical studies of one dimensional systems 
with arbitrary disorders and interactions. The argument is based on the 
study of the nodes of the wave functions, a simple exercise in one 
dimension which becomes highly nontrivial and still unsolved in higher 
dimensions. If the transverse dimensions are small, one can argue 
that nodal surfaces having a two dimensional topology will cost too much 
kinetic energy, and that the sign of $I$ is still given by the 
one dimensional rule. In the square lattices which we consider, 
we do not use Leggett's rule because the system is a narrow stripe, 
but because the Coulomb repulsion gives rise to a narrow stripe 
of current in the 2d lattice, reducing the two dimensional 
dynamics to a simpler one dimensional contribution. From the sign of 
$I$, one can know the parity of the effective number of particles 
which give rise to the observed one dimensional stripe of current.
 
\subsection{Disordered supersolid regime}
\label{subsection9.2}

 When we continue to decrease $U$, to eventually get a Fermi 
glass of three independent particles, we observe for $N=3$ a 
level crossing around $U_{lc}$ where the sign of $I$ changes. 
$I$ is diamagnetic above $U_{lc}$ and paramagnetic below $U_{lc}$. 
When $U$ is not too weak, the local currents forms a stripe of 
an almost one dimensional topology. This allows us to use Leggett's 
rule to argue that a diamagnetic current stripe is carried by $N=3$ 
particles, while the paramagnetic current stripe which persists below 
$U_{lc}$ is due to a single delocalized pair ($N=2$ even) in the 
background of an almost localized third particle. Below $U_{lc}$, 
the correlated pair gives a dominant paramagnetic current, while the 
third localized particle gives rise to a negligible $2d$ current of 
random sign. This disordered supersolid regime seems to be associated 
with the $1/W$-behavior of $I$ shown in Figs. \ref{Fig9} and \ref{Fig9a}. 

\begin{figure}[ht]  
\vskip.2in
\centerline{\epsfxsize=8cm\epsffile{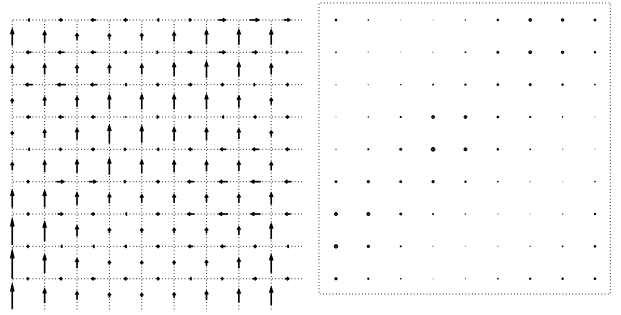}} 
\caption{Behavior obtained near the avoided level 
crossing ($U_{lc}$) for a disordered sample with $N=3$, 
$L=9$, $W=1$, $t=1$ and $U=15$. Left figure: map 
of currents for $\Phi=0.05 \Phi_0$. Right figure: 
corresponding site occupation numbers $n_{\bf j}^0$. 
The $n_{\bf j}^0$ are similar to those obtained 
in the same sample with $U=50$ (see Fig. \ref{Fig12}) 
while the currents is the superposition of two 
perpendicular diagonal motions. } 
\label{Fig12a}
\end{figure}

\begin{figure}[ht]  
\vskip.2in
\centerline{\epsfxsize=8cm\epsffile{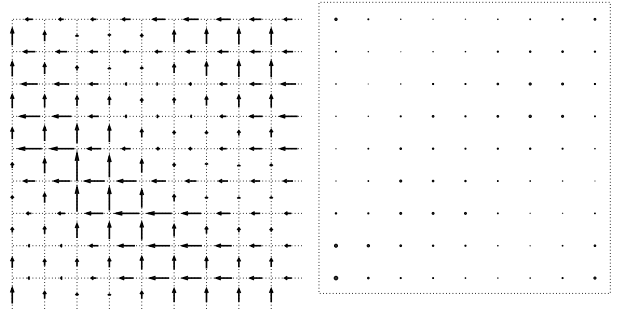}} 
\caption{Behavior obtained below the paramagnetic-diamagnetic 
crossover for a disordered sample with $N=3$, $L=9$, $W=1$, 
$t=1$ and $U=7$ ($r_s \approx r_s^F$) . 
Left figure: map of currents for $\Phi=0.05 \Phi_0$. 
Right figure: corresponding site occupation numbers $n_{\bf j}^0$. 
The left and right figures are now disconnected, $I$ flowing 
mainly perpendicularly to the axis where the $n_{\bf j}^0$ are maxima. 
} 
\label{Fig12b}
\end{figure}

Using the same sample which displays the continuum diamagnetic stripe 
shown in Fig. \ref{Fig12} for $U=50$, one shows the map of currents 
and the occupation numbers $n_{\bf i}^0$ in Fig. \ref{Fig12a} 
and  Fig. \ref{Fig12b} for $U \approx U_{lc}\approx  15$ and $U =7$ 
respectively. The  $n_{\bf i}^0$ are close to those obtained in the 
same sample for $U=50$ (see Fig. \ref{Fig12}). They form approximately 
the same stripe, though its width becomes broader. However, the map of 
currents are very different. As one can see in Fig. \ref{Fig12b} for 
$U=7$, the current stripe shown in the left figure is perpendicular to 
the density stripe visible in the right figure, while there is a 
superposition of two perpendicular diagonal flows near the avoided 
level crossing ( Fig. \ref{Fig12a}). The width of the paramagnetic 
stripe of current increases to give a two dimensional pattern of local 
currents when $U \rightarrow 0$ and $W/t \approx 1$, the sign of $I$ 
becoming sample dependent.

\subsection{$\Phi_0$, $\Phi_0/2$ and $\Phi_0/3$ harmonics of $I(\Phi)$}
\label{subsection9.4}

\begin{figure}[ht]  
\vskip.2in
\centerline{\epsfxsize=8cm\epsffile{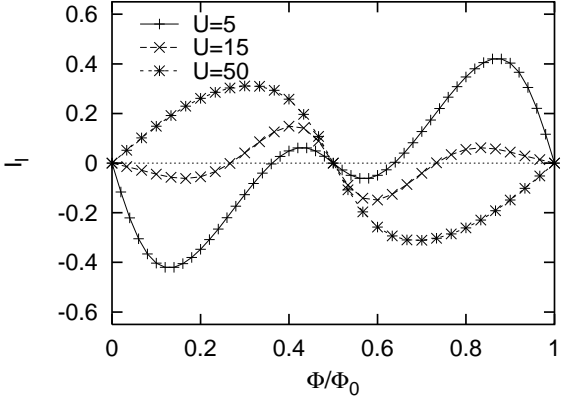}} 
\caption{$I_l$ as a function of $\Phi/\Phi_0$ 
for a disordered sample with $N=3$, $L=9$, $W=1$ and 
$t=1$ and three values of $U$. The negative (positive) 
currents are paramagnetic (diamagnetic) for an infinitesimal 
positive flux. One can see a change of sign between $U=15$ 
and $U=50$.} 
\label{Fig13}
\end{figure} 

 To study more precisely  how melts the Wigner molecule 
when $U \rightarrow 0$ in a disordered sample where 
$L=9$, $t=W=1$, we give in Fig. \ref{Fig13} $I_l$ as a function 
of $\Phi$, for the values of $U$ where one expects the end of the 
Fermi glass ($U=5$), a disordered supersolid ($U=15$) and the continuum 
CGSC-regime ($U=50$) respectively. For large $U$, Leggett's rule gives 
a diamagnetic molecule of $N=3$ particles, when the dynamics remains 
inside a 1d stripe. This rule is indeed observed for $U=50$ where 
$I_l(\Phi)$ exhibits an amplitude as large as for weaker interactions 
with no trace of other characteristic period than $\Phi_0$. In contrast, 
the curves obtained for $U=15$ and $U=5$ have reacher harmonic structures. 

$I_l(\Phi)$ is an odd function of $\Phi$ ($I_l(\Phi)=-I(\Phi)$) of 
period  $\Phi_0$ which can be expanded as:     
\begin{equation}
I_l(\Phi)= \sqrt{2} \sum_{n=1}^{\infty} I_n \sin \left( 2 \pi n
\frac{\Phi}{\Phi_0} \right)
\label{Fourier}
\end{equation}
where the Fourier components $I_n$ are 
\begin{equation}
I_n=\int_0^{\Phi_0} I_l(\Phi) \sqrt{2} \sin \left( 2 \pi n
\frac{\Phi}{\Phi_0} \right) d\Phi
\label{Fourier-component}
\end{equation}

\begin{figure}[ht]  
\vskip.2in
\centerline{\epsfxsize=8cm\epsffile{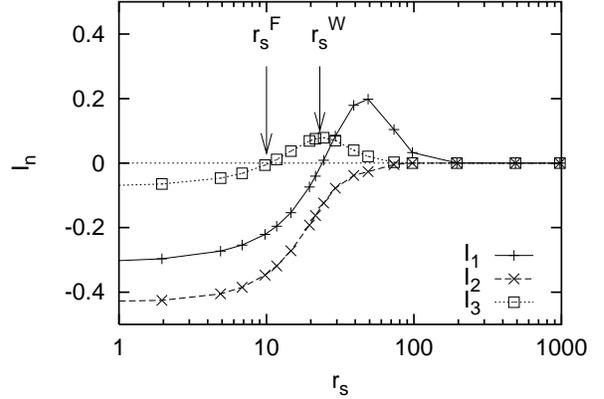}} 
\caption{Harmonics $I_n$ characterizing the period 
$\Phi_0/n$ of the longitudinal current $I_l(\Phi)$ 
as a function of $r_s$ for a disordered sample with 
$N=3$, $L=6$, $W=1$ and $t=1$.} 
\label{Fig14}
\end{figure}

\begin{figure}[ht]  
\vskip.2in
\centerline{\epsfxsize=8cm\epsffile{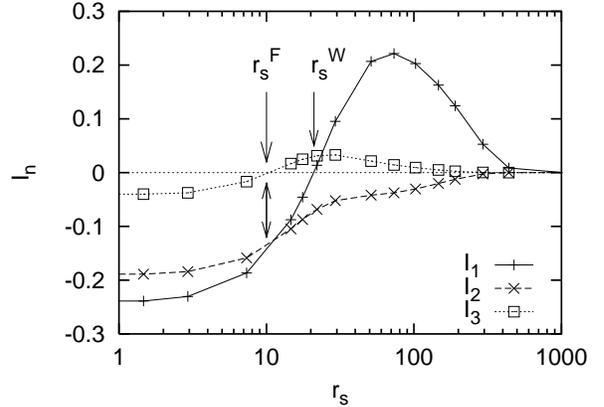}}  
\caption{Harmonics $I_n$ characterizing the period 
$\Phi_0/n$ of the longitudinal current $I_l(\Phi)$ 
as a function of $r_s$ for a disordered sample with 
$N=3$, $L=9$, $W=1$ and $t=1$.} 
\label{Fig14a}
\end{figure}

We give in Figs. \ref{Fig14} and \ref{Fig14a} the $\Phi_0$, $\Phi_0/2$ 
and $\Phi_0/3$ harmonics of $I_l(\Phi)$ as a function of $r_s$ for two 
disordered samples of size $L=6$ and $9$ respectively, with $W=t=1$. 
These harmonics, mainly $I_1$ and $I_2$, give the largest contributions. 
One can see that they are negative (paramagnetic) up to $r_s^F$. Around 
$r_s^F$, $I_3$ becomes positive, to reach a maximum for $r_s \approx 25$ 
where $I_1 \approx 0$. Above $r_s \approx 25$, the odd harmonics $I_1$ 
and $I_3$ are diamagnetic, while the even harmonic $I_2$ remain 
paramagnetic. This is what Leggett's rule gives if each $I_n$ was due 
to a $n$ body motion in one dimension. This allows us to give for each 
sample the characteristic values where the current topology becomes one 
dimensional and Leggett's rule applies. For $U\approx 40$ in the sample 
shown in Fig. \ref{Fig14a}, the diamagnetic $I_1$ is maximum while the 
two others are negligible, and one has the continuum stripe (CGS-regime) 
shown in  Fig. \ref{Fig12}. The period $\Phi_0$ should be expected for 
a continuum disordered stripe, a $L/3,L/3$ translation of each of the 
$3$ particles being equivalent to have a single particle enclosing $\Phi_0$.

\section{Summary}
\label{section10}

 Considering only $N=3$ spinless fermions in a 2d square lattice with 
random potentials, we have obtained many different regimes by increasing 
the strength $U$ of the electronic correlations. Studying how $I$ 
depends on the system parameters, we have distinguished the continuum 
regimes from the lattice regimes. Increasing $U$ in a disordered 
sample of size $L$, we have studied the effect of electronic correlations 
in the ballistic, diffusive and localized regimes without interaction. 
Looking at the map of local currents induced by a flux $\Phi$, we have 
identified two regimes reminiscent of phases discussed in other fields, 
the stripes in the theory of strongly correlated electrons and the 
supersolids in the theory of quantum solids.

 In the regime of the current stripe, the observed diagonal motion is 
clearly characteristic of the oblique shape of the Wigner molecule 
with $N=3$. For larger $N$, it is likely that one has a regime where 
the current flows along the axes of the Wigner crystal. If those axes 
form an angle with the shortest direction enclosing the flux, the 
ratio $I_t/I_l \neq 0$. When the electron crystal is pinned by disorder, 
this suggests that one can have an anisotropic resistivity tensor inside 
a certain range of density. Similar behaviors are observed in cuprate 
oxides with high temperature superconductivity in a certain range of 
chemical doping and in 2DEGs under high perpendicular magnetic fields. 
More directly related to our problem, the existence of stripe phases 
in 2DEGs is also discussed in Ref. \cite{spivak3}.  

 The nature of the intermediate regime between the solid and the liquid 
is a more delicate issue. We suggest that one could have a ``supersolid''. 
This concept was introduced long ago to describe the quantum melting of 3d 
quantum solids at very low temperature when the pressure is varied. Decades 
of studies of Helium-3 (fermions) or Helium-4 (bosons) atoms have not 
allowed to reach a firm conclusion. Very recent experiments by Kim and Chan 
in solid Helium-4 \cite{today,kim1,kim2,leggett2} can be interpreted as an 
observation of an apparent superfluid component, suggesting that a 
melting solid could have a supersolid fraction at certain intermediate 
pressures. For electrons in two dimensions, a supersolid phase is a 
possibility mentioned in Refs \cite{ksp,np}. One can also mention other 
possibilities. For instance a quantum liquid crystal \cite{bwp2} 
(quantum hexatic phase), in analogy with the thermal melting of an 
electron crystal of very large $r_s$. Clearly, more works are needed to 
understand more precisely the exact nature of the 2DEG at intermediate 
values of $r_s$. Let us conclude by summarizing three facts characterizing 
the intermediate values of $r_s$ : (i) the existence of an unexpected 
2d-metallic behavior given by transport measurements, (ii) 
the particular maps of local persistent currents obtained in exact numerical 
studies using a few spinless fermions, obtained in this work or in Refs 
\cite{bwp1,bwp2,avishai,sp,ksp,np}, (iii) the results of a recent fixed 
node Monte Carlo study \cite{fw} using $N \approx 50 - 200$ spinless 
fermions and extrapolated to the limit $N\rightarrow \infty$ which show 
that the nodal structure of a Slater determinant of delocalized Bloch waves 
gives a smaller GS energy than the trial wave-functions previously used to 
describe either a pure solid or a pure liquid, for intermediate values of 
$r_s$ ($30<r_s<80$). 

Z. \'A.\ N\'emeth \footnote{Present address: Department of Condensed 
Matter Physics, Weizmann Institute of Science, Rehovot 76100, Israel} 
acknowledges the financial support provided through the European 
Community's Human Potential Programme under contract HPRN-CT-2000-00144.

\end{document}